\documentclass[twocolumn,showpacs,preprintnumbers,amsmath,amssymb,APSl,prd,nofootinbib,superscriptaddress]{revtex4-2}

\usepackage{bm}
\usepackage{mathrsfs}
\usepackage{xcolor,color,graphicx,graphics}
\usepackage{colortbl}  
\usepackage{epsfig,subfigure}
\usepackage{latexsym,amssymb,amsmath,amsfonts} 
\usepackage[english]{babel} 
\usepackage[OT1]{fontenc}
\usepackage[latin1]{inputenc}
\usepackage{makeidx}
\usepackage{hyperref}
\usepackage{color,graphicx,graphics,wrapfig,epsf}
\usepackage{hyperref}
\hypersetup{colorlinks=true, linkcolor=blue, citecolor=green}
\usepackage{makecell}
\usepackage{multirow}
\def\imo{i}

\def\K{{\cal K}}
\def\Order#1{{\cal O}\left(#1\right)}
\usepackage{appendix}
\usepackage{orcidlink}

\usepackage{lipsum}     
\usepackage{mathtools}

\usepackage{latexsym}
\usepackage{orcidlink}
\usepackage{enumerate}

\definecolor{red}{rgb}{1,0,0}

\def\+{^\dagger}

\def\<{\leftarrow}
\def\>{\rightarrow}

\def\({\left(}
\def\){\right)}

\def\K{{\cal K}}

\newcommand{\bi}{\begin{itemize}} 				\newcommand{\ei}{\end{itemize}}
\newcommand{\benu}{\begin{enumerate}} 		\newcommand{\enu}{\end{enumerate}}
\newcommand{\bd}{\begin{dinglist}{0}}     \newcommand{\ed}{\end{dinglist}}
\newcommand{\bfig}{\begin{figure}[htbp]}  \newcommand{\efig}{\end{figure}}
        			
\newcommand{\bc}{\begin{center}} 				  \newcommand{\ec}{\end{center}}
\newcommand{\be}{\begin{equation}} 				\newcommand{\ee}{\end{equation}}
\newcommand{\bsub}{\begin{subequations}}  \newcommand{\esub}{\end{subequations}}
\newcommand{\ben}{\begin{eqnarray}} 			\newcommand{\een}{\end{eqnarray}}
\newcommand{\ba}[1]{\begin{array}{#1}} 		\newcommand{\ea}{\end{array}}
\newcommand{\bea}{\begin{equation}\begin{array}{rcl}}
\newcommand{\eea}{\end{array}\end{equation}}

\usepackage{soul}

\begin{document}
\title{Quasi-normal modes and shadows of scale-dependent regular black holes}

\author{Benjamin Koch
        \orcidlink{0000-0002-2616-0200}}
        \email{benjamin.koch@tuwien.ac.at}
        \affiliation{
Institut fur Theoretische Physik, TU Wien, Wiedner Hauptstrasse 8-10, 1040 Vienna, Austria, TU Wien, Atominstitut, Stadionallee 2, 1020 Vienna, Austria.}
\affiliation{Instituto de Fisica, Pontificia Catolica Universidad de Chile, Av. Vicu\~na Mackenna 4860, Santiago, Chile. }

\author{Gonzalo J. Olmo
        \orcidlink{0000-0001-9857-0412}}
        \email{gonzalo.olmo@uv.es}
        \affiliation{Departamento de F\'isica Te\'orica and IFIC, Centro Mixto Universitat de Valencia- CSIC. Universitat de Valencia, Burjassot-46100, Valencia, Spain.}
        \affiliation{Universidade Federal do Cear\'a (UFC), Departamento de F\'isica, Campus do Pici, Fortaleza- CE, C.P. 6030, 60455-760- Brazil.}

\author{Ali Riahinia
        \orcidlink{0000-0003-2736-1861}}
        \email{ali.riahinia@tuwien.ac.at}
        \affiliation{
Institut fur Theoretische Physik, TU Wien, Wiedner Hauptstrasse 8-10, 1040 Vienna, Austria, TU Wien, Atominstitut, Stadionallee 2, 1020 Vienna, Austria.}

\author{\'Angel Rinc{\'o}n 
        \orcidlink{0000-0001-8069-9162}} 
        \email{angel.rincon@physics.slu.cz}
        \affiliation{Research Centre for Theoretical Physics and Astrophysics, Institute of Physics, Silesian University in Opava, Bezru\v{c}ovo~n\'am\v{e}st\'i 13, CZ-74601 Opava, 
        Czech Republic.}
        \affiliation{Instituto Universitario de Matem\'atica Pura y Aplicada, Universitat Polit\`ecnica de Val\`encia, Valencia 46022, Spain.}

\author{Diego Rubiera-Garcia} \email{drubiera@ucm.es}
\affiliation{Departamento de F\'isica Te\'orica and IPARCOS, Universidad Complutense de Madrid, E-28040 Madrid, Spain.}

\date{\today}
\begin{abstract}

In this paper we investigate how a regular scale-dependent black hole, characterized by a single extra parameter $\epsilon$, behaves under perturbations by a test field (quasi-normal modes) and under light imaging (shadows) in a four-dimensional space-time background. On the quasi-normal modes side, we study how it responds to scalar and Dirac perturbations. To do this, we implement the well known WKB semi-analytic method of 6th order for obtaining the quasi-normal frequencies. We derive analytic expressions for the quasinormal frequencies beyond the eikonal limit for both scalar and Dirac perturbations, finding excellent agreement with the WKB approximation.
We discuss the behavior of the real and imaginary parts of the quasi-normal modes for different values of the parameter $\epsilon$ and the overtone $n$ and multipole $\ell$ numbers.
On the black hole imaging side, we ray-trace the geometry and illuminate it with a thin-accretion disk. Choosing $\epsilon=1.0$ we compute the size of the central brightness depression and generate full images of the black hole. We discuss the features (i.e. luminosity) of successive photon rings through the Lyapunov exponent of nearly-bound, unstable geodesics. Furthermore we use the correspondence (in the limit $\ell \gg n$) between quasi-normal mode frequencies and unstable bound light orbits to infer the numerical values of the latter using the former and find a remarkable accuracy of the correspondence in providing the right numbers. Our results support the usefulness of this correspondence in order to perform cross-tests of black holes using these two messengers.

\end{abstract}

\maketitle



\section{Introduction}\label{intro}


The search for a consistent theory of quantum gravity remains one of the major ongoing challenges in theoretical gravitational physics. One of the key requirements of such a theory is the resolution of space-time singularities. These correspond to the unavoidable development, anchored on Einstein's General Relativity (GR), of incomplete geodesic curves inside black holes and in the early stages of the cosmological evolution (see \cite{Senovilla:2014gza} for a discussion) and which typically go alongside the unbound growth of curvature scalars. This fact, which undermines the predictability of GR to describe the most extreme regions of the Universe, is at odds with the reliability of this very same theory to accurately describe the dynamics of black holes (via its Kerr solution). The latter statement is rooted on astrophysical X-ray tests \cite{Bambi:2017iyh}, gravitational wave observations of binary mergers \cite{LIGOScientific:2016aoc,LIGOScientific:2017vwq}, and imaging of the accretion flow around supermassive objects \cite{EventHorizonTelescope:2019dse,EventHorizonTelescope:2022wkp}. 

In view of the long-standing conceptual and technical difficulties in building such a theory of quantum gravity, approaches to remove or, at least, to ameliorate the degree of worry of such singularities, have led to the field of study of the so-called ``regular black holes" \cite{Bambi:2023try}. These geometries are aimed to either i) extend otherwise incomplete geodesic curves to new regions of space-time, ii) keep all curvature invariants finite everywhere, or iii) both of them.  Within this field, either such black holes are derived (analytically or numerically) from a well defined theory of gravity plus the matter fields, or a regularized {\it ad hoc}  line element is set first and then the action supporting it is re-constructed by driving back the corresponding field equations. Within any of these philosophies, these regular black hole scenarios are usually assumed to emerge from effective models incorporating quantum corrections and are believed to reflect, at least partially (e.g. from an effective field theory perspective \cite{Burgess:2003jk}), the regime where quantum gravitational effects become significant. Because of these reasons, regular black holes have become an attractive area of research to  studying theories beyond GR and to  get some glimpses of the long-sought quantum theory of gravity.

It is critical that any proposal for a regular black hole provides testable predictions for observable phenomena. Thus, beyond resolving the singularity problem, we must also investigate whether this regularization of the black hole interior leads to any observable consequences for the physics outside the event horizon.
In the context of black holes, two promising observables have recently received considerable attention in the literature: quasinormal modes (QNMs) and black hole imaging (or shadows). The QNMs, excited when a black hole is perturbed, encapsulate the characteristic ``ringdown'' signal and depend sensitively on the structure of the underlying space-time. 
Meanwhile, black hole shadows, now within reach of direct imaging techniques, offer a geometrical window into how light rays are bent and trapped near the photon sphere of unstable bound geodesics \cite{Falcke:1999pj,Perlick:2021aok}. Furthermore, it has been shown that QNMs (in the eikonal limit) and black hole imaging (in the strong deflection limit) quantities are directly related to each another~\cite{Stefanov:2010xz,Jusufi:2019ltj,Jusufi:2020dhz,Yang:2021zqy,Chen:2022nlw,Pedrotti:2024znu}. This enables an opportunity to perform cross-tests based on these two messengers.

The aim of this work is to study  a scale-dependent regular black hole in four space-time dimensions, where fundamental couplings such as Newton's constant vary with a scale parameter. This scale dependence aims to capture certain effective quantum corrections, thereby producing a black hole metric that remains regular at the origin. We then explore how these changes to the inner structure impact the QNM spectrum and the resultant black hole imaging. By combining insights from the regular nature of the black hole geometry with the rich phenomenology of QNMs and image formation, we hope to shed light on the subtle interplay between the quantum-corrected interior and the observable exterior. This study is driven by the need to understand how modifications to black hole interiors might impact external observables. It is thus important to bridge the expectations from theoretical models, motivated by effective quantum gravity-related considerations to cure classical pathologies, and the actual capabilities of experimental data (see e.g. \cite{Eichhorn:2022oma} for a discussion).

The paper is structured as follows: In Sec. \ref{S:II} we explain the framework of scale-dependent gravity. We review its motivation and links to asymptotic safety quantum gravity. We then show how the scale-dependent equations of motion are derived from the effective action. We briefly discuss various methods for promoting coupling constants to scale-dependent couplings and highlight the chosen method in this work. We also discuss the scale-dependent black hole solutions in this context. In Sec. \ref{S:III} we provide a brief explanation on QNMs and the Wentzel-Kramers-Brillouin (WKB) approximation used in our analysis. We study two cases: a spin zero case (scalar field) and a spin one-half case (Dirac field). We then explain how the WKB approximation method is used in our analysis and find the QNM frequencies for each case. In Sec. \ref{S:IV} we introduce the basic framework for black hole imaging, and produce simulated images of a thin-accretion disk for a case of a scale-invariant black holes satisfying recent constraints from the EHT. We pay special attention to the features of successive photon rings based on the employ of the Lyapunov exponent of nearly-bound unstable geodesics. In Sec. \ref{S:V} we discuss our results within the framework of the QNM-shadows correspondence, and illustrate it with the inference of quantities in the shadows' side found in Sec. \ref{S:IV}  from quantities of the QNM side found in Sec. \ref{S:V} to a great degree of precision. We conclude in Sec. \ref{S:VI} with a summary of our findings and some final thoughts.


\section{Background and formalism} \label{S:II}

 \subsection{Scale-dependent gravity}\label{sec2}
%
In this subsection, we aim to summarize the essential ingredients needed to understand the framework of scale-dependent gravity, a formalism strongly motivated by asymptotically safe gravity, one of several promising approaches to incorporating quantum features into gravity in a self-consistent manner.
The idea of formulating a complete (or at least better-defined) theory that unifies classical gravity with quantum mechanics is not new. It has had a profound and significative influence on the ongoing effort to develop a unified description of the universe \cite{Padmanabhan:2001ev,Padmanabhan:1998yy}.
There are well motivated physical scenarios in which the interplay between classical and quantum effects becomes particularly relevant. Black holes are a prime example. They provide an ideal setting that combines simplicity, observational accessibility, and deep physical insights, such as connections to statistical mechanics. These features make black holes an exceptional arena for exploring quantum aspects of gravity. In particular, within quantum-inspired gravitational theories, quantum corrections have been consistently incorporated into classical backgrounds, leading to fruitful developments.
There are, broadly speaking, three main approaches to introducing quantum corrections into classical gravitational solutions \cite{Reuter:2003ca,Rincon:2022hpy}:
\begin{enumerate}
    \item [i)] at the level of the solution,
    \item [ii)] at the level of the equations of motion, and
    \item [iii)] at the level of the action.
\end{enumerate}

An outstanding example of a modification of GR to include quantum effects is the pioneering work of Bonanno and Reuter \cite{Bonanno:2000ep}. Their work closely examines how renormalization group effects can perturb the classical Schwarzschild black hole solution, reshaping our understanding of its properties.
This method, often referred to as ``renormalization group improvement" of classical metrics or, more generally, improved black hole solutions, relies on a crucial concept: Newton's ``coupling constant" varies with scale. This running coupling is derived from the exact evolution equation for the effective average action and forms the core of their approach.

It is important to mention that such an improvement scheme does not provide actual solutions of the effective quantum-field equations and, therefore, it is not capable to capturing the full quantum nature of space-time. Nonetheless, it can provide valuable insights on the leading effects that mark the difference between a classical black hole and its quantum counterpart. Inspired by this breakthrough, a distinct line of research has flourished exploring further modifications to GR (see for example \cite{Koch:2014cqa, Cai:2010zh, Gonzalez:2015upa, Platania:2023srt, Ishibashi:2021kmf}). These studies continue to deepen our understanding of the interplay between quantum effects and gravitational theory.

Let us now turn to the central object that governs the dynamics in both classical and quantum gravitational theories: the action. In the presence of quantum corrections, the classical action is replaced by an effective action of the form $\Gamma[g_{\mu\nu}, k, \cdots]$, which incorporates the effects of quantum fluctuations. Here, $k \equiv k(x^{\mu})$ denotes an arbitrary energy (or renormalization) scale that may vary with the space-time position. 
In the case of static, spherically symmetric space-times, the scale dependence simplifies to $k(x^{\mu}) = k(r)$. This modification has a profound impact on the structure of the theory, as it promotes the coupling constants of the classical action (such as Newton's constant $G_0$ or the cosmological constant $\Lambda_0$) to scale-dependent functions. That is, these couplings become dynamical quantities that evolve with the scale $k$, or equivalently, with the radial coordinate. 

Since we shall begin our analysis with the effective action, the first step is to derive the corresponding effective Einstein field equations by varying $\Gamma[g_{\mu \nu}, k]$ with respect to the metric $g_{\mu \nu}$. A secondary equation arises from varying the action with respect to the scale field $k$. Although this procedure is, in principle, well defined, the resulting equation is highly complex and generally does not yield analytic solutions.
As an alternative, we adopt a simplified approach by imposing a saturated version of the null energy condition (NEC), as discussed in detail in the following sections. The combination of the effective Einstein equations and the NEC constraint allows us to obtain self-consistent solutions describing regular black holes within the framework of scale-dependent gravity.

It is worth highlighting a relevant fact of our formalism: the scale-dependent gravity framework is employed to establish a connection, and thereby identify, the effective stress-energy tensor with a running coupling constant, which can effectively generate regular black hole solutions. Consequently, although possible, we do not derive the metric functions directly from the equations of motion. At this stage, it becomes clear that we shall reinterpret these regular black hole solutions within the scale-dependent formalism. In particular, the lapse function adopted here belongs to a specific class of the Fan-Wang metric \cite{Fan:2016hvf}. As a result, the equations of motion of scale-dependent gravity are not utilized in the analysis; they are presented merely to ensure the paper is self-contained and to facilitate the identification of the relevant quantities in terms of scale-dependent gravity.

\subsection{Scale-dependent regular black hole}

To derive the effective field equations, we begin by considering the following scale-dependent effective action \cite{Contreras:2017eza}:
\begin{equation}
\Gamma[g_{\mu\nu},k] = \int d^4x \sqrt{-g} \left[\frac{1}{2\kappa_k}(R - 2\Lambda_k)\right] + S_k,
\end{equation}
where the symbols retain their conventional meanings: 
i) $R$ is the Ricci scalar, 
ii) $\Lambda_k$ is the scale-dependent cosmological parameter, and 
iii) $S_k$ represents the matter action. We work in units where $c = 1$ 
and $\kappa \equiv 8\pi G$.
As previously discussed, all classical couplings are promoted to scale-dependent functions. Varying the action with respect to the metric yields the effective Einstein field equations:
\begin{equation}
G_{\mu\nu} + \Lambda_k g_{\mu\nu} = \kappa_k (T^{\textrm{eff}})_{\mu\nu},
\label{einstein}
\end{equation}
where $(T^{\textrm{eff}})_{\mu\nu}$ denotes the effective energy-momentum tensor, defined as
\begin{equation}
(T^{\textrm{eff}})_{\mu\nu} \equiv (T_{\mu\nu})_k - \frac{1}{\kappa_k} \Delta t_{\mu\nu}.
\end{equation}
The auxiliary tensor $\Delta t_{\mu\nu}$ arises from the scale dependence of Newton's constant and is defined as
\begin{equation}
\Delta t_{\mu\nu} = G_k \left( g_{\mu\nu}\,\square - \nabla_\mu \nabla_\nu \right) G_k^{-1}.
\end{equation}
To determine the scale-setting function $k(x)$, we vary the effective action with respect to $k$ \cite{Koch:2014joa}, requiring minimal dependence:
\begin{align}\label{scale}
\frac{\mathrm{d}}{\mathrm{d}k} \Gamma[g_{\mu \nu}, k] = 0.
\end{align}

Combining the effective Einstein equations \eqref{einstein} with the scale-setting condition \eqref{scale} closes the system, allowing for the construction of self-consistent solutions. Inserting such a solution into $\Gamma_k$, up to boundary terms, yields an effective action compatible with the underlying symmetries.
However, due to the lack of a precise form for the effective action $\Gamma_k$, particularly the beta functions governing quantum gravitational couplings, such a procedure remains incomplete in practice. As previously discussed, an additional and widely adopted closure condition involves imposing the saturated version of the NEC. Applying the NEC to the effective energy-momentum tensor leads to
\begin{equation}
(T^{\text{eff}})_{\mu\nu} \ell^{\mu} \ell^{\nu} = \left[ T_{\mu\nu} - \frac{1}{\kappa(r)} \Delta t_{\mu\nu} \right] \ell^{\mu} \ell^{\nu} \geq 0,
\end{equation}
where $\ell^\mu$ is a null vector.

Assuming a static and spherically symmetric spacetime, the line element takes the form  
\begin{equation}\label{m1}
\mathrm{d}s^{2} = -A(r)\,\mathrm{d}t^{2} + B(r)\,\mathrm{d}r^{2} + r^2\,\mathrm{d}\Omega^2,
\end{equation}  
where $\text{d}\Omega^2 \equiv \text{d}\theta^2 + \sin^2\theta\,\text{d}\phi^2$. For simplicity, we consider the case $\{A(r) = B(r)^{-1} \}$. A convenient choice for a radial null vector is $(\ell^{\mu} = \{f^{-1/2}, f^{1/2}, 0, 0\})$. Imposing the NEC on the effective energy-momentum tensor, 
\begin{equation}
(T^{\text{eff}})_{\mu\nu} \ell^{\mu} \ell^{\nu} = 0,
\end{equation}  
and noting that ($G_{\mu\nu} \ell^{\mu} \ell^{\nu} = 0$), consistency with the effective field equations \eqref{einstein} requires that  
\begin{equation} \label{necnm}
\Delta t_{\mu\nu} \ell^{\mu} \ell^{\nu} = 0.
\end{equation}  
Solving Eq.~\eqref{necnm} yields the explicit form of the running Newton's coupling \cite{Rincon:2017ayr},  
\begin{equation} \label{gr}
G(r) = \frac{G_0}{1 + \epsilon r},
\end{equation}  
where $G_0$ is the classical Newton constant and $\epsilon$ is the scale-dependent parameter with dimensions of inverse length. As expected, setting $\epsilon = 0$ recovers the classical solution, while $\epsilon \neq 0$ introduces quantum corrections.
The idea and formalism have been applied in several fields, for instance in  
  i)  black hole physics (and their properties) 
 ii)  wormholes 
iii) relativistic stars and
 iv) cosmological models (see e.g. Refs. 
\cite{Rincon:2017goj,Rincon:2018sgd,Rincon:2018lyd,Rincon:2019zxk,Fathi:2019jid,Contreras:2018gpl,Rincon:2020cpz,Rincon:2019cix,Sendra:2018vux,Panotopoulos:2021tkk,Contreras:2018swc,Canales:2018tbn,Alvarez:2022wef,Panotopoulos:2021heb,Alvarez:2022mlf,Alvarez:2020xmk,Bargueno:2021nuc,Panotopoulos:2021obe,Panotopoulos:2020zqa}).

It is essential to mention that the scale-dependent scenario generally leads to results that differ from those obtained within the framework of asymptotically safe gravity (see Fig.~4 in Ref.~\cite{Rincon:2019cix} and related discussion). Asymptotic safety is one of the most conventional and well established frameworks aiming to describe quantum gravity consistently. However, these differences can be reconciled by considering the so-called ``infrared instability'' (see Ref.~\cite{Biemans:2016rvp}), which further motivates the investigation of black hole space-times within the scale-dependent gravity formalism.
To further reinforce the viability of this approach, we highlight the fact that scale-dependent gravity has also been successfully applied in a cosmological context. Specifically, it has been used to: i) address the cosmological constant problem, ii) alleviate the tension in the Hubble constant $H_0$, and iii) investigate the discrepancies between the standard $\Lambda$ CDM model and observational data regarding the rms amplitude of matter fluctuations, $\sigma_8$. For detailed discussions on these topics, see e.g. Refs.~\cite{Alvarez:2020xmk,Bargueno:2021nuc,Panotopoulos:2021heb}.

In a series of papers Donoghue and collaborators emphasized the distinction between mathematical cut-off running and physical running of gravitational couplings
\cite{Donoghue:2019clr,Donoghue:2024uay}. As a consequence,
they question the validity of a universal running of e.g. the gravitational coupling $G_k$. However, they perfectly acknowledge the validity of a contextual~\footnote{By contextual we mean that the apparent running will depend on the effective theory AND on the observable which is chosen to test this theory} scale dependence of this coupling, say for a black hole, where e.g. $G\rightarrow G(r)$~\cite{Donoghue:1994dn,Anber:2010uj,Anber:2011ut,Donoghue:1993eb,Bjerrum-Bohr:2002gqz}. Therefore, in what follows we will focus on a contextual scale-dependence $G=G(r)$, constructed to frame a regular black hole in the language of scale-dependence. This specific black hole previously analyzed in~\cite{Contreras:2017eza}, is characterized by the metric functions  
\begin{equation} \label{metricelement}
A(r) = B(r)^{-1} = 1 - \frac{2MG_0}{r} \left(1 + \frac{M^2 G_0^2 \epsilon}{6r}\right)^{-3}
\end{equation}
where $M$ denotes the black hole mass, $G_0$ is Newton's gravitational constant in four dimensions, and $\epsilon > 0$ is a scale-dependent parameter with dimensions of inverse length.
This line element is regular everywhere and exhibits a de Sitter core in the limit $r \to 0$, where the lapse function takes the form  
\begin{align}
A(r) \to 1 - \frac{432\, r^2}{G_0^5 M^5 \epsilon^3}.
\end{align}
From this expression, we can identify an effective cosmological constant, in term of the scale-dependent parameter $\epsilon$, given by
\begin{align}
\Lambda_{\textrm{eff}} = \frac{1296}{G_0^5 M^5 \epsilon^3}.
\end{align}

This regular solution can be interpreted without the inclusion of a cosmological constant, instead attributing its shape to the presence of matter in an anisotropic vacuum, which modifies the Schwarzschild geometry. In the $\epsilon \rightarrow 0$ limit, the Schwarzschild solution is recovered. Since the running parameter $\epsilon$ modulates quantum effects, it is expected to take small values.
Moreover, in the asymptotic limit $r \rightarrow \infty$ an additional correction term of the form $\epsilon M^3 / r^2$ appears. This far-field behaviour corresponds to the correction found in Ref.~\cite{Scardigli:2014qka} (see Eq.~29 of that work), giving us the identification $\tilde{\epsilon} = M \epsilon$.
For practical purposes, the metric given in Eq.~\eqref{metricelement} can be recast in terms of dimensionless variables by introducing $T = t/M$, $x = r/M$, and $\tilde{\epsilon} = \epsilon M$. Setting $G_0 = 1$, the line element becomes \cite{Sendra:2018vux}
\begin{equation}\label{m2}
\mathrm{d}s^{2} = -A(x)\,\mathrm{d}T^{2} + B(x)\,\mathrm{d}x^{2} + x^2\,\mathrm{d}\Omega^2,
\end{equation}
with the metric functions given by
\begin{equation}
A(x) = B(x)^{-1} = 1 - \frac{2}{x} \left(1 + \frac{\tilde{\epsilon}}{6x}\right)^{-3},
\label{metricelement2}
\end{equation}
The event horizon, which defines the limit in which trapped surfaces are developed, can be obtained simply by finding the roots of $A(x) = 0$. This leads to the cubic equation:
\begin{equation}
216\,x^3 - (432 - 108\,\tilde{\epsilon})\,x^2 + 18\,\tilde{\epsilon}^2 x + \tilde{\epsilon}^3 = 0.
\label{xh}
\end{equation}
so that the largest real root of this equation, $x_H = \max\{x_1, x_2, x_3\}$, corresponds to the location of the event horizon.

\section{Quasinormal modes}\label{S:III}

In what follows, we present the key equations and fundamental theory required to introduce the reader to the calculation of QNMs. Our focus is on the modified Schwarzschild black hole introduced in the previous section, and reinterpreted in the context of scale-dependent gravity. We consider massless scalar and Dirac field perturbations and use the well-established Wentzel-Kramers-Brillouin (WKB) approximation up to sixth order to determine the corresponding quasinormal frequencies.

While we shall adopt one of the most well established semi-analytical methods for QNM computations, it is worth mentioning that there is a variety of analytical and numerical techniques that have been developed in the literature, each with its own pros and cons.
The choice of method typically depends on the complexity of the space-time geometry and the nature of the perturbing fields. However, it is worth noting that exact analytical solutions for black hole QNM spectra are only known for a limited class of space-times. Notable examples are:
i) P\"oschl-Teller Potential: when the effective potential barrier takes the form of a P\"oschl-Teller potential, as explored in \cite{Poschl:1933zz, Ferrari:1984zz, Cardoso:2001hn, Cardoso:2003sw, Molina:2003ff, Panotopoulos:2018hua}, QNM frequencies can be obtained through a simple analytical relation (see subsequent paragraphs);
ii) Hypergeometric Functions: When the radial wave equation can be transformed into Gauss's hypergeometric equation, as discussed in \cite{Birmingham:2001hc, Fernando:2003ai, Fernando:2008hb, Gonzalez:2010vv, Destounis:2018utr, Ovgun:2018gwt, Rincon:2018ktz}, 
iii) Heun Functions: in certain cases, such as Teukolsky's equations for Kerr-de Sitter black holes, the governing equations can be mapped onto Heun's equations, allowing for an exact determination of QNM frequencies using Heun functions \cite{Hatsuda:2020sbn, Fiziev:2011mm, Naderi:2024dhh}.

These analytical methods provide valuable insights into QNM spectra and serve as important benchmarks for numerical approaches. Due to the complexity and non-trivial nature of the differential equations involved, the calculation of QNM frequencies often requires numerical or semi-analytical methods. Several techniques have been developed for this purpose, including:
i) Frobenius Method and its generalizations \cite{Destounis:2020pjk, Fontana:2022whx, Hatsuda:2021gtn},
ii) Continued Fraction Method and its refinements \cite{Leaver:1985ax, Nollert:1993zz, Daghigh:2022uws},
iii) Asymptotic Iteration Method \cite{Cho:2011sf, 2003JPhA...3611807C, Ciftci:2005xn}.
For a more comprehensive review of these and other methods, see \cite{Konoplya:2011qq}. Finally, it is worth emphasizing that the quasinormal modes of regular black holes in asymptotically safe gravity (and in closely related scenarios) have already been calculated with high precision in previous works (see, e.g., Refs.~\cite{Konoplya:2022hll,Konoplya:2023aph}).

\subsection{Spin zero case (scalar field)}

Let us consider the dynamics of a test scalar field, $\Phi$, propagating in a four-dimensional space-time background. In the following we shall assume a real scalar field. We thus take the corresponding action, $S[g_{\mu \nu}, \Phi]$, 
\begin{align}
S[g_{\mu \nu} ,\Phi] \equiv \frac{1}{2} \int \mathrm{d}^4 x \sqrt{-g}
\Bigl[
\partial^{\mu} \Phi \partial_{\mu} \Phi 
\Bigl]\,.
\end{align}
Based on such an action, we obtain the corresponding equation of motion for the massless scalar field \cite{Crispino:2013pya,Kanti:2014dxa,Pappas:2016ovo,Panotopoulos:2019gtn,Avalos:2023ywb,Gonzalez:2022ote,Rincon:2020cos}
\begin{equation}
\frac{1}{\sqrt{-g}}\partial_{\mu}\left(\sqrt{-g}g^{\mu\nu}\partial_{\nu}\Phi\right) = 0.
\end{equation}
Taking advantage of the symmetries of the metric, we can use a convenient way to decouple the perturbations. Let us assume an angular space spanned by spherical harmonics as eigenfunctions, which is achieved by the sum over all possible eigenvalues,
\begin{equation}
\Phi(t, r, \theta, \phi) 
=\sum_{\ell ,m}e^{-i\omega t}\frac{\psi(r)}{r}Y_{\ell m}(\theta, \phi).\label{fdc}
\end{equation}

It is important to mention that we adopt the standard monochromatic ansatz with harmonic time dependence $e^{-i\omega t}$, corresponding to a single Fourier mode of the scalar field perturbation (with fixed multipole indices $\ell, m$). This is sufficient for analyzing the quasinormal modes and linear evolution, as arbitrary perturbations can be decomposed into such modes due to the linearity of the equations.

After applying the above ansatz the differential equation can be written in the form
\begin{align}
\begin{split}
& \frac{\omega^{2}r^{2}}{A(r)} + \frac{r}{\psi(r)}\frac{\mathrm{d}}{\mathrm{d}r}\left[r^{2}A(r)\frac{\mathrm{d}}{\mathrm{d}r}\left(\frac{\psi(r)}{r}\right)\right] = \ell(\ell+1)
\label{KG}
\end{split}
\end{align}
where we have used the angular part and replace this part in terms of its eigenvalues, i.e., 
\begin{align}
    \begin{split}
&\frac{1}{\sin\theta}\frac{\partial}{\partial\theta}\left(\sin\theta\frac{\partial Y}{\partial\theta}\right) + \frac{1}{\sin^{2}\theta}\frac{\partial^{2}Y}{\partial\phi^{2}} = 
-\ell(\ell + 1)Y,
\label{kg2}
\end{split}
\end{align}
with $Y \equiv Y(\Omega) = Y(\theta, \phi)$, while  $\ell(\ell + 1)$ is the eigenvalue, and $\ell$ is the angular degree. As it can be seen the  combination of Eqs. (\ref{KG}) and (\ref{kg2}) generate a second-order differential equation for the radial coordinate. To rewrite the resulting equation in a more convenient way, it becomes essential to  introduce the so-called ``tortoise coordinate" $r_{*}$, defined as follows
\begin{equation}
\label{tcd1}
    \mathrm{d}r_{*}  \equiv \frac{\mathrm{d}r}{A(r)}\,, 
\end{equation}
which induces a rule of transformation for the derivative, $A(r) \partial_r = \partial_{r_*}$. Taking the last change of variable into account, we rewrite the resulting differential equation on its canonical Schr{\"o}dinger-like form
\begin{equation} \label{SLE}
\frac{\mathrm{d}^{2}\psi(r_*)}{\mathrm{d}r_{*}^{2}} + \left[\omega^{2} - V(r)\right]\psi(r_*) = 0,
\end{equation}
where $V(r)$ is the effective potential defined via the expression 
\begin{equation}
V(r) = A(r)
\Bigg[ 
\frac{\ell(\ell + 1)}{r^{2}} + \frac{A'(r)}{r}
\Bigg],\label{poten}
\end{equation}
and where the prime represents the derivative of the radial variable with respect to the tortoise coordinate.

Finally, to fully determine the QNM frequencies, imposing suitable boundary conditions is needed, which generally depend on the asymptotic behavior of the space-time. In the present case, the relevant conditions are:
\begin{align}
  \Phi \rightarrow \: &\exp(-i \omega r_*), \; \; \; \; \; \;  r_* \rightarrow - \infty ,
   \\
   \Phi \rightarrow \: &\exp(+i \omega r_*), \; \; \; \; \; \; r_* \rightarrow + \infty .
\label{pbc}
\end{align}
for purely ingoing modes at the event horizon and purely outgoing at asymptotic infinity, respectively. Assuming a time dependence of the form $\Phi \sim e^{-i \omega t}$, a negative imaginary part of the frequency $\omega$ corresponds to a decaying mode, indicating stability under perturbations. Conversely, a positive imaginary part leads to a growing mode, implying instability. Thus, the stability of the black hole against scalar perturbations is determined by the sign of the imaginary part of the QNM frequency.

\begin{figure*}[ht!]
	\centering
	\includegraphics[scale=0.618]{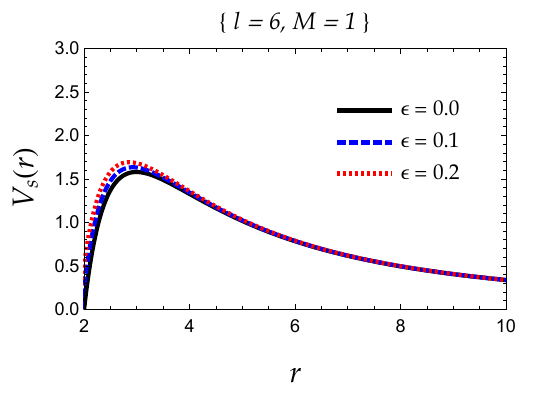} \
	\includegraphics[scale=0.618]{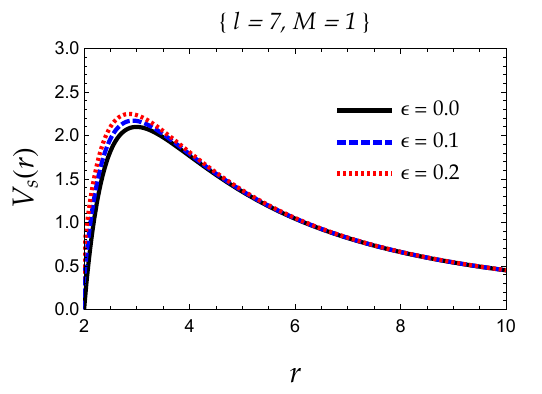} \  
        \includegraphics[scale=0.618]{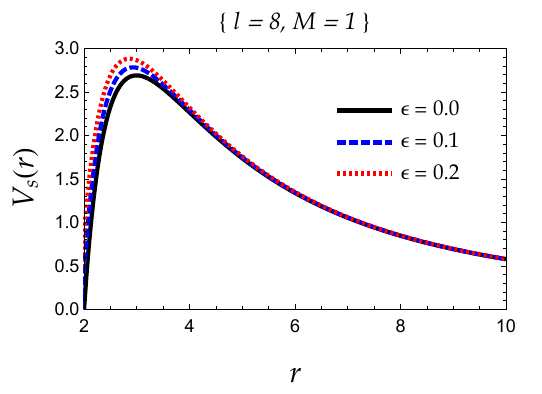} 
        \\
        \includegraphics[scale=0.618]{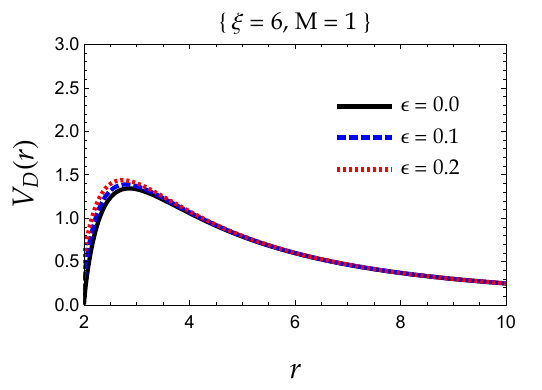} \
	\includegraphics[scale=0.618]{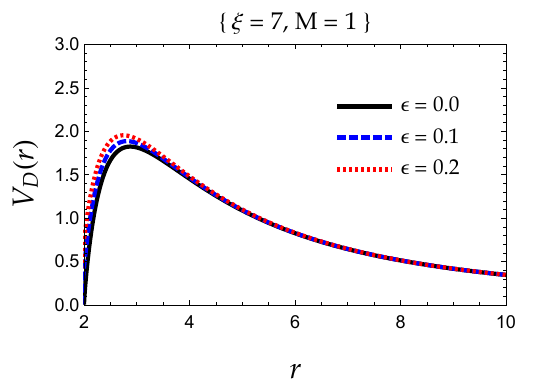} \  
        \includegraphics[scale=0.618]{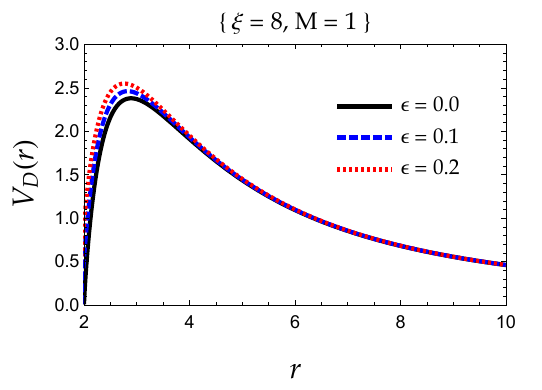} 
    \caption{
		Scalar and Dirac effective potentials for different values of the parameters $\{M, \epsilon, \xi \text{ (or } \ell)\}$.
		{\bf{First Row:}} Scalar effective potential for a fixed mass $M=1$ varying the parameters $\ell$ and $\epsilon$.
	{\bf{Second Row:}} Dirac effective potential for a        fixed mass $M=1$ varying the parameters $\xi$ and         $\epsilon$.
	}
	\label{fig:1} 	
\end{figure*}

In the first row of Fig.~\ref{fig:1} we show the shape of the effective potential barrier $V_s(r)$ for scalar perturbations against the radial coordinate $r$ for different values of the parameter set $\{ \epsilon, \ell \}$ for a fixed value of the black hole mass $M$. 
We observe that when the scale-dependent parameter $\epsilon$ is different from zero, we slightly modify the Schwarzschild black hole solution by taking quantum features into account.
Since we have only one free parameter in addition to the black hole mass (in this case, the scale-dependent parameter $\epsilon$), we vary this number for different values of the angular number $\ell$. 
We notice that as $\epsilon$ increases, the maximum of the potential also increases, at the time shifting to the left, which means that the strength of the effective potential is slightly higher than in the GR solution. Furthermore, all the curves converge for large radii, while there are small differences for small radii. The same behaviour is repeated when we increase $\ell$, making the difference between the classical case and its corresponding quantum-corrected version more pronounced.

\subsection{Spin one-half case (Dirac field)}

In this subsection we shall present the basic framework for obtaining the corresponding equation of motion of the neutral Dirac particles and then calculate the QNMs of such a field. We restrict our considerations for the same spherically symmetric background considered in the previous section.
We begin by performing a conformal transformation in the line element as (see \cite{Das:1996we,Gibbons:1993hg} and references therein)
\begin{eqnarray}
g_{\mu\nu} & \rightarrow & \overline{g}_{\mu\nu}=\tilde{\Omega}^{2}g_{\mu\nu} , 
\\
\psi & \rightarrow & \overline{\psi}=\tilde{\Omega}^{-3/2}\psi , 
\\
\gamma^{\mu}\nabla_{\mu}\psi & \rightarrow & \tilde{\Omega}^{5/2} \overline{\gamma}^{\mu}\overline{\nabla}_{\mu}\overline{\psi} ,
\end{eqnarray}
with $\overline{\psi}=r^{3/2}\psi$ and the conformal factor, $\tilde{\Omega}=1/r$. The line element  is then cast in the following form
\begin{equation}
\label{metric2}
\mathrm{d}\overline{s}^{2} = -\frac{1}{r^{2}}A(r)\mathrm{d}t^{2} + \frac{1}{r^{2}}A(r)^{-1}\mathrm{d}r^{2} + \mathrm{d}\Omega^{2}. 
\end{equation}
Now we can separate the $(t,r)$ part from the two-spheres sector, which allows us to decouple the equation to study the radial-temporal part. Thus, we rewrite the massless Dirac perturbation in the form $\overline{\gamma}^{\mu}\overline{\nabla}_{\mu}\overline{\psi} = 0$, i.e,
\begin{align}
\begin{split}
\label{diraceq}
\bigg[ 
\left(
\overline{\gamma}^{t}\overline{\nabla}_{t} +
\overline{\gamma}^{r}\overline{\nabla}_{r} \right) \otimes 1
+ 
\overline{\gamma}^{5} \otimes
\left( \overline{\gamma}^{a}
\overline{\nabla}_{a}\right)_{S_{2}} 
\bigg] 
\overline{\psi} 
&= 0 , 
\end{split}
\end{align}
where $\overline{\gamma}^{\mu}, \mu=0,1,2,3$ are Dirac's gamma matrices, while $\overline{\gamma}^5=i \overline{\gamma}^0 \overline{\gamma}^1 \overline{\gamma}^2 \overline{\gamma}^3$ with $(\overline{\gamma}^{5})^{2}=1$, and the field ansatz to be considered is
\begin{equation}
\overline{\psi} = \sum_{\ell} 
\left( 
\phi_{\ell}^{(+)}  \chi_{\ell}^{(+)}  + 
\phi_{\ell}^{(-)}  \chi_{\ell}^{(-)}  
\right) \ ,
\end{equation}
where $\phi_{\ell}^{(\pm)} \equiv \phi_{\ell}^{(\pm)} (r,t)$ and $\chi_{\ell}^{(\pm)} \equiv \chi_{\ell}^{(\pm)} (\theta , \phi )$.
We use the fact that the angular part admits fixed eigenvalues $\pm i(\ell + 1)$, which we denote by $\pm i \xi$, and consider $\chi_{\ell}^{(\pm)}$ as the eigenspinors on the two-spheres (for further details, see \cite{Camporesi:1995fb}), that is
\begin{equation}
\left( \gamma^{a}\nabla_{a} \right)_{S_2}\chi_{\ell}^{(\pm)} = \pm i \left( \ell + 1\right) \chi_{\ell}^{(\pm)}.
\end{equation}
Notice that $\ell = 0, 1, 2, \dots$ and Eq.(\ref{diraceq}) is rewritten (for notational simplicity, from now on we drop the bars) as follows
\begin{equation}
\label{eqn:2Ddirac}
\left \{ 
\gamma^{t} \nabla_{t} + \gamma^{r} \nabla_{r} + \gamma^{5} \left[ \pm i \left( \ell + 1 \right) \right] \right \} 
\phi_{\ell}^{(\pm)} = 0 .
\end{equation}
The last equation is the pair of two-dimensional Dirac equations in the coordinates $r$ and $t$. 
To solve the differential equations, we make an explicit choice of the Dirac matrices, namely
\begin{align}
\gamma^{t} &= \frac{r}{\sqrt{A(r)}}(-i\sigma^{3}), 
\\
\gamma^{r} &=  \sqrt{A(r)} 
\ r\sigma^{2} .
\end{align}
Here $\sigma^{i}$ are the well-known Pauli matrices, defined as
\begin{equation} \nonumber
\sigma^{1}=\left(
\begin{array}{cc}
0 & 1 \\ 1 & 0
\end{array}
\right)\ \ \ ,\ \ \ \sigma^{2}=\left(
\begin{array}{cc}
0 & -i \\ i & 0
\end{array}
\right)\ \ \ ,\ \ \ \sigma^{3}=\left(
\begin{array}{cc}
1 & 0 \\ 0 & -1
\end{array}
\right) .
\end{equation}
As usual, $\gamma^{5}$ can be written by using the Pauli matrices
\begin{equation}
\gamma^{5} = (-i\sigma^{3})(\sigma^{2}) = - \sigma^{1} .
\end{equation}
The spin connections are then found to be:
\begin{align}
\Gamma_{t} &= \sigma^{1}  
\frac{r^2}{4} \frac{\mathrm{d}}{\mathrm{d}r} \left( \frac{A(r)}{r^{2}} \right) , \\
\Gamma_{r} &= 0 .
\end{align}
Since the treatment for $\phi_{\ell}^{(-)}$ is equivalent to $\phi_{\ell}^{(+)}$, we will concern ourselves only with the positive value in the sequel. Also, we have to set the spinor in terms of two different components, according to the rule
\begin{equation}
\phi_{\ell}^{(+)} = \left( \frac{\sqrt{A(r)}}{r} \right)^{-1/2} e^{-i \omega t} \left(
\begin{array}{c}
i\mathcal{G} (r) \\ \mathcal{F} (r)
\end{array}
\right) ,
\end{equation}
and then replacing it into the Dirac equation we obtain
\begin{equation}
\begin{split}
&\sigma^{2} \left(  \sqrt{A(r)} \ r \right) \left(
\begin{array}{c}
i\frac{\mathrm{d} \mathcal{G} (r)}{\mathrm{d}r} \\ 
\ \frac{\mathrm{d} \mathcal{F} (r)}{\mathrm{d}r}
\end{array}
\right) -i \sigma^{1} \xi \left(
\begin{array}{c}
i \mathcal{G} (r) 
\\ 
\mathcal{F} (r)
\end{array}
\right) = 
\\
&\sigma^{3} \omega \left( \frac{r}{\sqrt{A(r)}} \right) \left(
\begin{array}{c}
i \mathcal{G} (r) 
\\ 
\mathcal{F} (r)
\end{array}
\right) .
\end{split}
\end{equation}
By treating the components separately, we finally write a set of coupled first-order differential equations in terms of the variables $\mathcal{G}  \equiv \mathcal{G} (r)$ and $\mathcal{F} \equiv \mathcal{F} (r)$ as follows
\begin{eqnarray}
A(r) \frac{\mathrm{d} \mathcal{G} (r)}{\mathrm{d}r} 
- 
\Bigg [ \frac{\sqrt{A(r)}}{r} \xi \Bigg ] \mathcal{G} (r) & = & + \omega \mathcal{F} (r)   \label{Eq1}
\\ 
A(r) \frac{\mathrm{d} \mathcal{F} (r)}{\mathrm{d}r} 
+ 
\Bigg [ \frac{\sqrt{A(r)}}{r} \xi \Bigg ] \mathcal{F} (r) & = & - \omega \mathcal{G} (r) . \label{Eq2}
\end{eqnarray}
Now we introduce the special potential $\mathcal{W} \equiv \xi \sqrt{A(r)}/r$, in terms of which the Dirac potentials will be suitably rewritten later on. Considering this potential and the tortoise coordinate defined in (\ref{tcd1}), we rewrite the pair of equations (\ref{Eq1})-(\ref{Eq2}) as
\begin{align}
    \Bigg[
    \frac{\mathrm{d}}{\mathrm{d}r_{*}} - \mathcal{W} 
    \Bigg] \mathcal{G} &= + \omega\mathcal{F} ,
    \\
    \Bigg[
    \frac{\mathrm{d}}{\mathrm{d}r_{*}} + \mathcal{W} 
    \Bigg] \mathcal{F} &= - \omega \mathcal{G}.
\end{align}
This set of first-order differential equations for $\mathcal{G}$ and $\mathcal{F}$ can be easily decoupled to obtain two Schr\"odinger-like differential equations, with two concrete effective potentials, i.e.,
\begin{align}
\frac{\mathrm{d}^2\mathcal{F}}{\mathrm{d}{r_{*}}^2} + [\omega^2 - V_{-}] \mathcal{F} & =  0 , \label{SL1}
\\
\frac{\mathrm{d}^2 \mathcal{G}}{\mathrm{d}{r_{*}}^2} + [\omega^2 - V_{+}] \mathcal{G} & =  0 , \label{SL2}
\end{align}
where the potentials are given by 
\begin{equation}
V_{\pm}(r) = \mathcal{W}^2 \pm \frac{\mathrm{d} \mathcal{W} }{\mathrm{d}r_{*}} = \mathcal{W}^2 \pm A(r) \frac{\mathrm{d} \mathcal{W} }{\mathrm{d}r}.
\end{equation}
In general, the effective potentials $V_{+}$ and $V_{-}$ are expected to exhibit supersymmetric properties with respect to each other, generating identical QNM spectra for the functions $\mathcal{G}$ and $\mathcal{F}$ (i.e., isospectral potentials). This equivalence typically holds for both wave scattering and QNM analysis. However, there exist space-time geometries where this isospectrality is broken, providing physical motivation to study the deviation. In what follows, we present the explicit form of both potentials for the space-time configurations under consideration, namely
\begin{equation}
\label{potex}
V_\pm = A(r) 
\Bigg(
\frac{\xi^2}{r^2} \pm \xi 
    \Bigg[
        \frac{A'(r)}{2r \sqrt{A(r)}} - \frac{\sqrt{A(r)}}{r^2}
    \Bigg]
\Bigg).
\end{equation}
In the following, we will consider only the positive sign of the effective potential for simplicity. The analysis for the negative sign is entirely analogous, so we restrict our attention to this single case. Accordingly, we will denote $V_D \equiv V_{+}$ from now on.
At this point we should choose appropriate outgoing boundary conditions at the horizon and at spatial infinity. Usually for Schr\"{o}dinger-like equations like those of \eqref{SL1} and \eqref{SL2} the possible physically relevant boundary condition is expressed as
\begin{align}
   \{\mathcal{F}, \mathcal{G}\} \rightarrow \: &\exp(-i \omega r_*), \; \; \; \; \; \;  r_* \rightarrow - \infty ,
   \\
   \{\mathcal{F}, \mathcal{G}\} \rightarrow \: &\exp(+i \omega r_*), \; \; \; \; \; \; r_* \rightarrow + \infty .
\end{align}
The last condition can be different depending on the analyzed space-time and field potential at the asymptotic region.

In the second row of Fig.~\ref{fig:1} we show the form of the effective potential barrier $V_D(r)$ for Dirac perturbations against the radial coordinate $r$ for different values of the set of parameters $\{ \epsilon, \xi \}$ for a fixed value of the black hole mass $M$. 
As in the scalar case, when the scale-dependent parameter is shifted from zero to positive values, the effective potential is altered (softly) as a result of the quantum modifications.
To be more precise, the figures confirm the same pathology observed in the scalar case, i.e. as $\epsilon$ increases, the maximum of the effective potential increases, at the same time as it shifts to the left. The same behaviour is also observed when we increase $\xi$, highlighting the difference between the classical and the scale-dependent counterpart.

%

\subsection{Method: the WKB approximation}

As mentioned above, the effective potential is well behaved, which is the reason why we will utilize the WKB semi-classical method to compute the corresponding QNM frequencies (see \cite{Schutz:1985km,Iyer:1986np,Iyer:1986nq,Kokkotas:1988fm,Seidel:1989bp,Lutfuoglu:2025hjy,Hamil:2024njs,Hamil:2024ppj} and references therein). 
Originally introduced by Schutz and Will at first order \cite{Schutz:1985km}, further extended by Iyer and Will to include second and third-order corrections \cite{Iyer:1986np}, significant improvements were made by R.~A.~Konoplya, who extended the method to the sixth order \cite{Konoplya:2003ii}, and by Matyjasek and Opala, who extended it to the 13th order \cite{Matyjasek:2017psv}, improving its precision and applicability to a larger class of potentials. 
This method (particularly in its lower-order formulations) has proven effective in calculating the fundamental QNM of black holes, especially for the Schwarzschild case. Its accuracy improves with increasing angular harmonic index $\ell$ (with $\ell \propto \xi$), but decreases for higher overtone numbers.

It is worth noting that the convergence of higher order WKB approximations has not been established mathematically. As a heuristic argument, Konoplya proposed to estimate errors by comparing the QNM frequencies of successive WKB orders. Although useful, this approach does not provide a rigorous criterion for selecting the optimal order.
In practice, the sixth or seventh order tends to give the most accurate results, although this depends sensitively on the background geometry. Therefore, no single order can be universally preferred.
The technique is based on solving a one-dimensional Schr\"odinger-like equation with a potential barrier, and involves fitting asymptotic solutions by a Taylor expansion around the peak of the potential at $x = x_0$. This expansion is valid in the region between the turning points, defined as the roots of $U(x, \omega) \equiv V(x) - \omega^2$.

In this paper, we apply the WKB method to compute QNM  frequencies up to the sixth order, using the generalized formula:
\begin{equation}
\omega_n^2 = V_0 + \sqrt{-2V_0''}\, \Lambda(n) - i\, \nu \sqrt{-2V_0''} \left[1 + \Omega(n)\right], 
\end{equation} 
where:  
(i) $V_0$ is the peak of the effective potential, 
(ii) $V_0''$ denotes its second derivative at the peak, 
(iii) $\nu = n + 1/2$ with $n=0,1,2,\dots$ the overtone number, and (iv) $\Lambda(n)$ and $\Omega(n)$ are correction functions explicitly given in \cite{Kokkotas:1988fm}.  
We use a Wolfram Mathematica \cite{wolfram} implementation of the WKB method, which supports orders one to six \cite{Konoplya:2019hlu}. In our analysis, we restrict to the regime $n < \xi \text{ (or } \ell)$ to ensure the validity of the approximation. For higher-order corrections, we refer the reader to \cite{Konoplya:2019hlu,Hatsuda:2019eoj}.

\begin{figure*}[t!]
	\centering
	\includegraphics[scale=0.934]{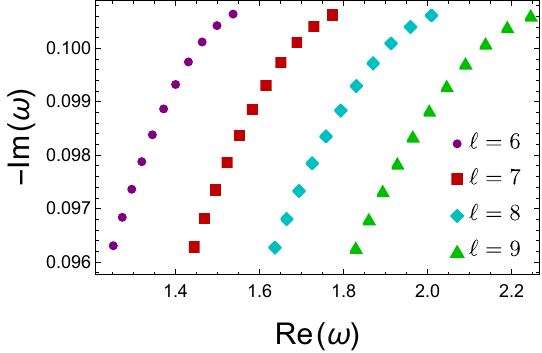} \
	\includegraphics[scale=0.934]{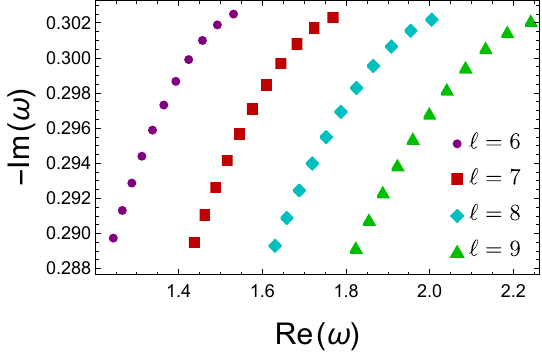} \  
    \caption{
		QNMs for massless scalar perturbations.
		{\bf{Left Panel:}} QNMs for $M=1$, varying $\epsilon$ from 0.0 to 1.0 for different values of the parameter $\ell$ for the fundamental mode $(n=0)$.
		{\bf{Right Panel:}} QNMs for $M=1$, varying $\epsilon$ from 0.0 to 1.0 for different values of the parameter $\ell$ for the first excited mode $(n=1)$.
	}
	\label{fig:2} 	
\end{figure*}

\begin{figure*}[t!]
	\centering
	\includegraphics[scale=0.934]{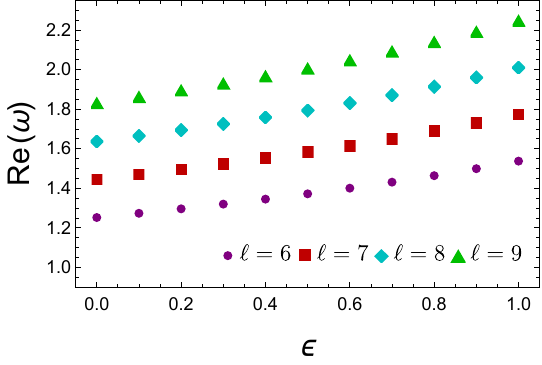} \
	\includegraphics[scale=0.934]{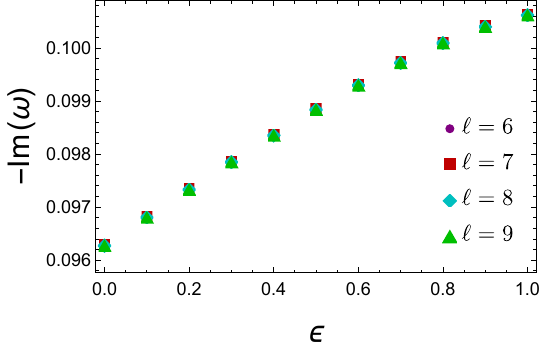} \  
    \\
        \includegraphics[scale=0.934]{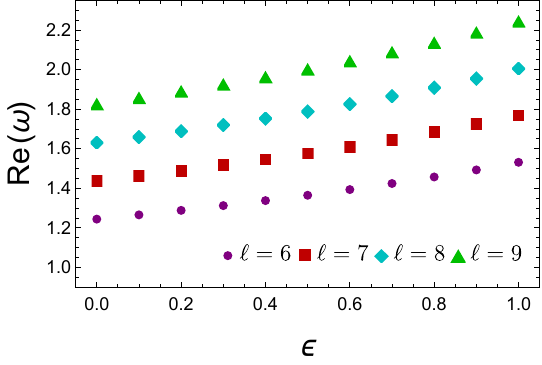} \
	\includegraphics[scale=0.934]{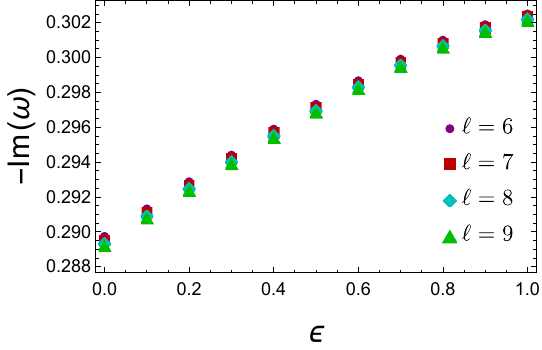} \  
    \caption{Real/imaginary part of scalar QNM frequencies, $\omega_R$/$\omega_{I}$, against the scale-dependent parameter $\epsilon$.
		{\bf{Top Left Panel:}} $\text{Re}(\omega)$ vs $\epsilon$ for $M=1$, varying $\ell$ from 6 to 9  for the fundamental mode $(n=0)$.
		{\bf{Top Right Panel:}} $-\text{Im}(\omega)$ vs $\epsilon$ for $M=1$, varying $\ell$ from 6 to 9  for the fundamental mode $(n=0)$.
        {\bf{Down Left Panel:}} $\text{Re}(\omega)$ vs $\epsilon$ for $M=1$, varying $\ell$ from 6 to 9 for the first exited mode $(n=1)$.
		{\bf{Down Right Panel:}} $-\text{Im}(\omega)$ vs $\epsilon$ for $M=1$, varying $\ell$ from 6 to 9 for the first excited mode $(n=1)$.
	}
	\label{fig:4} 	
\end{figure*}

\begin{figure*}[t!]
	\centering
	\includegraphics[scale=0.934]{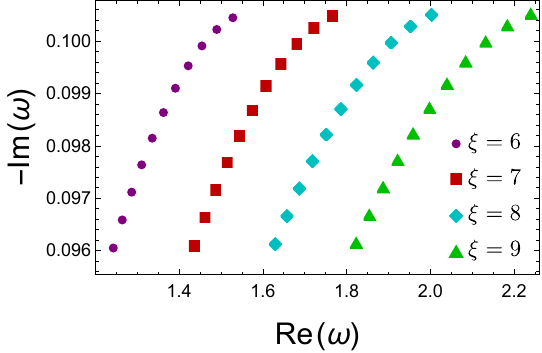} \
	\includegraphics[scale=0.934]{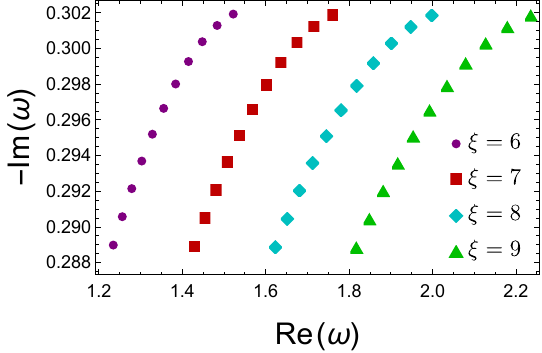} \  
    \caption{
		QNMs for massless Dirac perturbations.
		{\bf{Left Panel:}} QNMs for $M=1$, varying $\epsilon$ from 0.0 to 1.0 for different values of the parameter $\xi$ for the fundamental mode $(n=0)$.
		{\bf{Right Panel:}} QNMs for $M=1$, varying $\epsilon$ from 0.0 to 1.0 for different values of the parameter $\xi$ for the first exited mode $(n=1)$.
	}
	\label{fig:3} 	
\end{figure*}

\begin{figure*}[t!]
	\centering
	\includegraphics[scale=0.934]{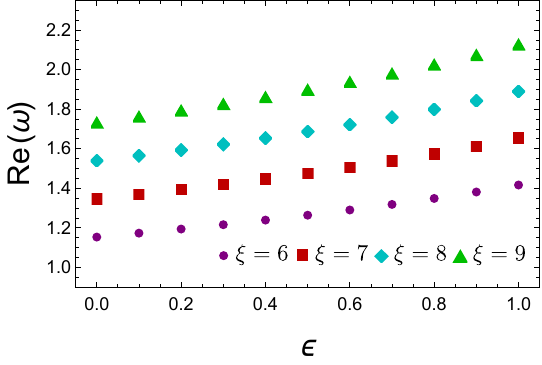} \
	\includegraphics[scale=0.934]{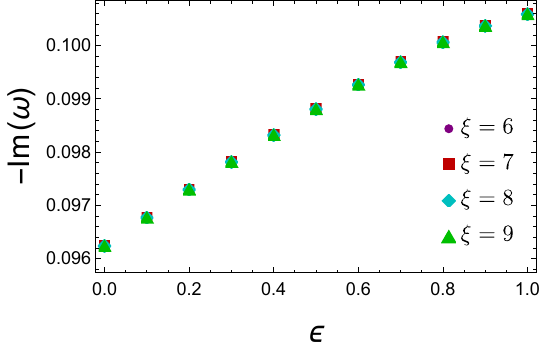} \  
    \\
        \includegraphics[scale=0.934]{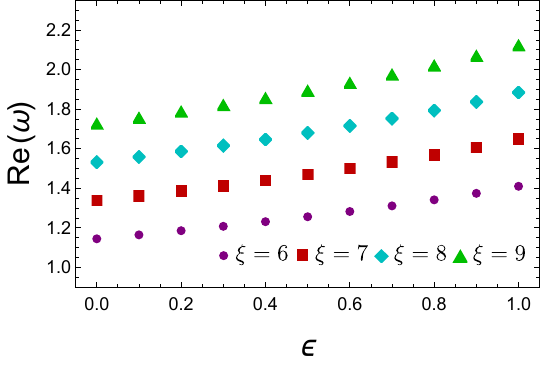} \
	\includegraphics[scale=0.934]{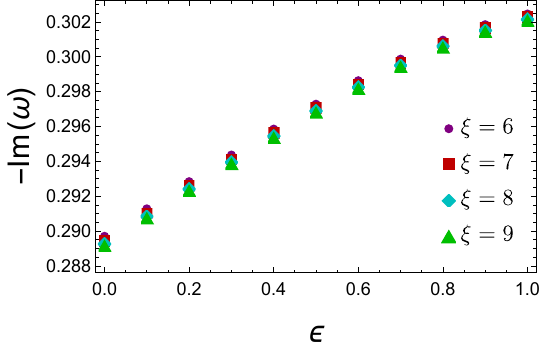} \  
    \caption{Real/imaginary part of Dirac quasinormal frequencies, $\omega_R$/$\omega_{I}$, against the scale-dependent parameter $\epsilon$.
		{\bf{Top Left Panel:}} $\text{Re}(\omega)$ vs $\epsilon$ for $M=1$, varying $\xi$ from 6 to 9  for the fundamental mode $(n=0)$.
		{\bf{Top Right Panel:}} $-\text{Im}(\omega)$ vs $\epsilon$ for $M=1$, varying $\xi$ from 6 to 9  for the fundamental mode $(n=0)$.
        {\bf{Down Left Panel:}} $\text{Re}(\omega)$ vs $\epsilon$ for $M=1$, varying $\xi$ from 6 to 9 for the first exited mode $(n=1)$.
		{\bf{Down Right Panel:}} $-\text{Im}(\omega)$ vs $\epsilon$ for $M=1$, varying $\xi$ from 6 to 9 for the first exited mode $(n=1)$.
	}
	\label{fig:5} 	
\end{figure*}

\subsection{On analytic expressions for QNMs}

Following the spirit of the WKB approximation, and remembering that the WKB series converges only asymptotically (and also ensure the exact result in the eikonal regime only), recently it was noticed by Konoplya and Zhidenko \cite{Konoplya:2023moy} 
a non-trivial way to obtain analytical expressions for the QNMs (and grey-body factors) of black holes, by expanding beyond the eikonal approximation. We have then followed closely this idea and formalism to supplement our numerical result obtaining a very good agreement with our findings. 
This idea has been applied satisfactorily in some papers,  see for instance \cite{Malik:2024voy,Malik:2024sxv,Malik:2024tuf} and references therein.
For the sake of this paper, let us focus on the underlying idea and critical expressions. 
The effective potential can be approximated by the expression:
\begin{equation}\label{potential-multipole}
V(r_*)=\kappa^2\Bigl(H(r_*)+\Order{\kappa^{-1}}\Bigl),
\end{equation}
where $\kappa\equiv\ell+1/2$. 
Once the form of the effective potential has been established, we can expand the position of its maximum and the frequency in terms of $\kappa^{-1}$. However, for our specific model, this expression is inconvenient. For the purposes of this paper, therefore, we will consider only lower-order corrections (with respect to $\epsilon$) to the maxima of the effective potential.
Also, since the function $H(r_*)$ solely has a maximum, its position (see Eq.(\ref{potential-multipole})) can be expressed as follows:
\begin{equation}\label{rmax}
  r_{\max }=r_0+r_1\kappa^{-1}+r_2\kappa^{-2} + \cdots .
\end{equation}
For illustration, let us use (\ref{rmax}) into the following first order WKB formula for the frequency
\begin{eqnarray}
\omega&=&\sqrt{V_0-\imo \K\sqrt{-2V_2}},
\end{eqnarray}
and subsequently expanding it in terms of powers of $\kappa^{-1}$, it is possible to obtain \cite{Konoplya:2023moy}
\begin{eqnarray}\label{eikonal-formulas}
\omega=\Omega\kappa-\imo\lambda\K+\Order{\kappa^{-1}}.
\end{eqnarray}
At this point, it is clear that: 
 i) $\Omega$ represents the angular velocity at the unstable null geodesics, and 
ii) $\lambda$ is the Lyapunov exponent. 
In addition, $\K \equiv \nu = n+1/2$.
The last expression can naturally be compared with \eqref{eikonal-formulas}, which was obtained by Cardoso et al \cite{Cardoso:2008bp}. They linked unstable null geodesics around a static, spherically symmetric, asymptotically flat or de Sitter black hole with its quasinormal frequencies in the regime where $\ell \gg n$. 
The expression is written via the equation \cite{Cardoso:2008bp}:
\begin{equation}\label{QNM}
\omega_n=\Omega\ell-\imo(n+1/2)|\lambda|, \quad \ell \rightarrow \infty.
\end{equation}
Subsequently, in \cite{Konoplya:2017wot} the scenarios in which such a correspondence 
is violated are discussed, including for instance the case of Einstein-(dilaton)-Gauss-Bonnet and Einstein-Lovelock theories \cite{Konoplya:2017wot,Konoplya:2020bxa,Konoplya:2019hml}). 
Even more, for asymptotically de Sitter BHs the overtones number $n$ is not the frequency number \cite{Konoplya:2022gjp}. The latter is true because the quasinormal spectrum consists of two distinct branches: the Schwarzschild branch and the de Sitter branch. The de Sitter branch cannot be reproduced using the standard WKB approach (which is valid only in the eikonal regime dominated by the black hole geometry). Consequently, the analytical continuation procedure that works so well for the Schwarzschild branch is not applicable to the de Sitter branch.

The analytic expression for the scalar case, considering up to third order in powers of $\epsilon$ and the third order beyond eikonal limit is written as follow:
\begin{widetext}
\begin{align} \label{wanaliticscalar}
\begin{split}
\omega = 
&
\hspace{0.4cm}
  \frac{29}{1296 \sqrt{3} \kappa  M}
- \frac{23 \K^2}{108 \sqrt{3} \kappa  M} 
+ \frac{\kappa }{3 \sqrt{3} M}
- \frac{i \K}{3 \sqrt{3} M}
+
\\
& \hspace{0.4cm}
\epsilon
\Bigg(
\frac{35 \K^2}{1944 \sqrt{3} \kappa }+\frac{31}{23328 \sqrt{3} \kappa }+\frac{\kappa }{18 \sqrt{3}}-\frac{i \K}{54 \sqrt{3}}
\Bigg)
\
+
\\
& \hspace{0.4cm}
\epsilon^{2}
\Bigg(
\frac{185 \K^2 M}{23328 \sqrt{3} \kappa }-\frac{131 M}{279936 \sqrt{3} \kappa }+\frac{\kappa  M}{72 \sqrt{3}}+\frac{i \K M}{1944 \sqrt{3}}
\Bigg)
+
\\
& \hspace{0.4cm}
\epsilon^{3}
\Bigg(
\frac{1225 \K^2 M^2}{419904 \sqrt{3} \kappa }-\frac{1747 M^2}{5038848 \sqrt{3} \kappa }+\frac{49 \kappa  M^2}{11664 \sqrt{3}}+\frac{47 i \K M^2}{34992 \sqrt{3}}
\Bigg)
\
+
\Order{\epsilon^4,\frac{1}{\kappa^4}}
\end{split}
\end{align}
where we have used the expansion in powers of $\kappa^{-1}$, similarly to \cite{Konoplya:2023moy}. Thus, we have obtained the expansion for the location of the potential maximum for a massless scalar field, namely:
\begin{equation}\label{rmax-scalar}
\begin{array}{rcl}
r_{\max } &=& \displaystyle3 M-\frac{M}{3 \kappa ^2}
+
\epsilon  
\Bigg(
\frac{5 M^2}{27 \kappa ^2}-\frac{2 M^2}{3}
\Bigg)
+
\epsilon^2
\Bigg(
\frac{M^3}{162 \kappa ^2}-\frac{M^3}{18}
\Bigg)
+
\epsilon^3
\Bigg(
\frac{M^4}{486 \kappa ^2}-\frac{7 M^4}{486}
\Bigg)
+
\mathcal{O}\left(\epsilon ^4,\frac{1}{\kappa ^4}\right).
\end{array}
\end{equation}

Similarly, for the Dirac field, the corresponding QNM frequency, at first order in $\epsilon$ and lower expansion beyond eikonal limit, is given by 
\begin{align}\label{eikonal-Dirac}
\begin{split}
\omega  =
&
\hspace{0.4cm} 
\frac{\kappa }{3 \sqrt{3} M} 
- 
\frac{i \K}{3 \sqrt{3} M} 
+
\frac{23 \K^2}{216 \sqrt{3} \kappa ^2 M} 
+ 
\frac{7}{2592 \sqrt{3} \kappa ^2 M} 
- 
\frac{23 \K^2}{108 \sqrt{3} \kappa  M} 
- 
\frac{7}{1296 \sqrt{3} \kappa  M} 
+ 
\frac{1}{6 \sqrt{3} M}
\\
&
\hspace{0.4cm}
\epsilon
\Bigg(
\frac{5}{46656 \sqrt{3} \kappa ^2} 
-
\frac{35 \K^2}{3888 \sqrt{3} \kappa ^2}
+
\frac{35 \K^2}{1944 \sqrt{3} \kappa }
-
\frac{5}{23328 \sqrt{3} \kappa }
+
\frac{\kappa }{18 \sqrt{3}}
-
\frac{i \K}{54 \sqrt{3}}
+
\frac{1}{36 \sqrt{3}}
\Bigg)
+
\mathcal{O}\left(\epsilon ^2,\frac{1}{\kappa ^4}\right),
\end{split}
\end{align}
while the position of the maximum of the potential is
\begin{equation}\label{rmax-Dirac}
r_{\max } = 
 3 M - \frac{\sqrt{3} M}{2 \kappa } + \frac{\sqrt{3} M}{4 \kappa ^2} 
+
\epsilon
\Bigg(
\left(4-5 \sqrt{3}\right)\frac{M^2}{72 \kappa ^2}+\frac{5 M^2}{12 \sqrt{3} \kappa }-\frac{2 M^2}{3}
\Bigg)
+
\mathcal{O}\left(\epsilon ^2,\frac{1}{\kappa ^4}\right).
\end{equation}
\end{widetext}
As seen in Tables~\ref{tab:3}--\ref{tab:4}, the frequencies of the fundamental mode show excellent agreement with those obtained using the direct WKB semi-analytical method whenever $\ell > 0$ or $\xi > 0$ (see also the purple- and olive-highlighted rows in Tables~\ref{tab:1}--\ref{tab:2} in the appendix for a detailed comparison). Furthermore, the following points should be highlighted:
i) In the scalar case, both the real and imaginary parts of the frequency show excellent agreement with the WKB results as the quantum parameter $\epsilon$ is increased (up to moderately large values). Consistent with the WKB approximation, $\mathrm{Re}(\omega)$ increases with $\epsilon$, while $-|\mathrm{Im}(\omega)|$ (i.e., the decay rate) also grows with increasing $\epsilon$.
ii) In the Dirac case, the behavior is qualitatively similar to the scalar case, though minor discrepancies appear. The real part of the QNM frequency remains in good agreement with the WKB prediction; the imaginary part agrees better. However, both components of QNM frequencies exhibits slightly larger deviations when $\epsilon$ increases. These small discrepancies are expected, since our analysis goes only to leading order beyond the eikonal limit and is restricted to first order in $\epsilon$ for simplicity. Consequently, as $\epsilon$ becomes larger, some deviation from the WKB results (and other alternative approaches) is naturally anticipated.

Tables \ref{tab:5} and \ref{tab:6} show the percentage error between the WKB method and the analytic expressions beyond the eikonal approximation.
We have used the standard definition of percent error, i.e.,  
\begin{align} \label{ErrorR}
    \delta_R(\omega_R) &\equiv \Bigg| \frac{\omega_R^{\text{WKB}} - \ \omega_R^{\text{analytic}}}{\omega_R^{\text{analytic}}}\Bigg|\times100\% ,
    \\
    \label{ErrorI}
      \delta_R(\omega_I) &\equiv \Bigg| \frac{ \omega_I^{\text{WKB}} - \ \omega_I^{\text{analytic}} }{ \omega_I^{\text{analytic}} }\Bigg|\times100\%,
\end{align}
where we also define the error, component by component, using
\begin{align}\label{Error}
    \delta_R &\equiv (\delta_R(\omega_R) , \delta_R(\omega_I)).
\end{align}
From the tables, we can confirm several key points:
i) For scalar perturbations, the error in the real part of the quasinormal frequencies reaches a maximum of approximately $0.3\%$, occurring for large values of $\epsilon$ and small values of $\ell$, as expected. Thus, the real part is essentially identical between the two methods, meaning that the oscillation frequency of the perturbation is insensitive to the choice of approach. Additionally, the error in the imaginary part reaches a maximum of $\sim 0.08\%$, again for large $\epsilon$ and small $\ell$. In both the real and imaginary parts, the error decreases as $\ell$ increases, consistent with the improved performance of the WKB approximation at higher multipoles.
ii) For Dirac perturbations, the situation is largely similar (with $\ell$ replaced by $\xi$), although the errors are somewhat larger than in the massless scalar case. The error in the real part reaches a maximum of $\sim 1.5\%$ for small $\xi$ and large $\epsilon$, decreasing as $\xi$ increases. The imaginary part exhibits a maximum error of $\sim 0.1\%$ for large values of both $\epsilon$ and $\xi$. These slightly larger errors are expected, given the successive approximations employed in the computation (including those for the lapse function, the location of the potential maximum, and the expansion of $\kappa$) which inevitably introduce a minor loss of precision. Nevertheless, even with these intermediate approximations, the errors remain well controlled, as confirmed by Table \ref{tab:6}.
In summary, we have demonstrated numerically that both the WKB and analytic methods yield highly accurate results, at least for the parameter values considered in this work.

\subsection{Discussion of results}

We have summarized our findings in Figs. \ref{fig:2} and \ref{fig:4} for scalar perturbations and in Figs. \ref{fig:3} and \ref{fig:5} for Dirac perturbations. Furthermore, we have also collected all the numerical values obtained of the computation of the real and imaginary parts of the QNMs in Tables \ref{tab:1} and \ref{tab:2} in the Appendix \ref{A:QNM}. 
For comparison, we have derived analytical expressions for the QNM frequencies in both the scalar and Dirac cases (see Eqs.~\eqref{wanaliticscalar}-\eqref{eikonal-Dirac}). The modes computed from these expressions are summarized in Tables~\ref{tab:3} and~\ref{tab:4} of Appendix~\ref{A:QNM} and show very good agreement with the WKB approximation.
In particular, we observe, in Fig. \ref{fig:2}, the QNM spectrum for $M=1$ varying the scale-dependent parameter $\epsilon$ (implicitly). We plot the fundamental mode ($n=0$ left panel) and the first exited mode ($n=1$ right panel), for different values of the angular number $\ell$. For a fixed $\ell$ we observe that the modes increase and move to the right.
Then, in Fig. \ref{fig:4} we show the response of the real and imaginary QNM frequencies versus the scale-dependent parameter. We clearly observe that when $\epsilon$ increases, the modes depart from the standard Schwarzschild solution, regardless of the $\ell$ value. In addition, the figures confirm that the real part of the QNM is more sensitive than the imaginary part when the scale-dependent parameter is varied.

It is worth pointing out that  the period of oscillation for the QNM is determined by the value of the $\text{Re}(\omega_n)$ term as $T \propto  2 \pi/\text{Re}(\omega_n)$. More precisely, this is the frequency component of the ringdown observed after the merger of compact objects, such as black holes. As the figures show, in this simplified model, a potential observational imprint emerges naturally via the alteration of the real part of the QNM frequencies. Similarly, the damping time of the QNM is determined by the value of the $\text{Im}(\omega_n)$ term, provided that such an imaginary part is negative (otherwise it indicates instability of the system).

On the other hand, Figs. \ref{fig:3} and \ref{fig:5} show the response of the real and imaginary QNM frequencies against the scale-dependent parameter, in this case, for the Dirac field. According to our numerical results, the behavior is equivalent to that observed in the scalar case: the period of oscillation is affected when $\epsilon$ is non-zero and so is the damping time.

\section{Shadows} \label{S:IV}

In this section we consider the problem of the optical appearance of our modified Schwarzschild black hole of scale-invariant gravity when illuminated by a thin-accretion disk, a field known as black hole imaging but, more popularly, simply as {\it shadows}. This is of special relevance under the light of the recent developments in very-long baseline interferometry allowing the detection of such optical appearance for supermassive black holes, as proven by the images of the central objects within the M87 \cite{EventHorizonTelescope:2019dse} and Milky Way galaxies 
\cite{EventHorizonTelescope:2022wkp}. 

The theoretical basis leading to black hole imaging revolves around light (null) trajectories driven by the radial and angular potentials of geodesic motion. While highly complicated in the axially symmetric (e.g. Kerr) case, great simplifications are achieved in the present spherically symmetric case allowing, to a great extent, for an analytical approach to this problem (see e.g. \cite{Perlick:2021aok} for a thorough discussion). Indeed, the equations for null motion can be cast, in the present case, as 
\begin{equation}
\left(\frac{dr}{d\lambda}\right)^2=\frac{1}{b^2}-V(r)
\end{equation}
where $\lambda$ is the affine parameter, the impact parameter $b \equiv L/E$ is just the ratio between the (conserved) particle's angular momentum and energy, while the potential $V(r)$ reads simply as
\begin{equation} \label{eq:V(r)s}
    V(r)=\frac{A(r)}{r^2}
\end{equation}
Photon trajectories arriving to the observer's screen and whose impact parameter equals the critical one, the latter given by the maxima of the potential, $b=b_c$, that is

\begin{align}
    V(r)\vert_{r=r_{ps}} &=\frac{1}{b^2} ; \nonumber
    \\
    V'(r)\vert_{r=r_{ps}} &=0 ;
    \\
    V''(r)\vert_{r=r_{ps}} &<0 \nonumber
\end{align}
asymptote, when traced backwards, to the unstable surface of bound geodesics $r=r_{ps}$ (defined by the first of the equations above, while the other two state the fact that it corresponds to a maximum of the potential), usually dubbed as the {\it photon sphere}. Both the critical impact parameter $b_c$ and the photon sphere $r_{ps}$ can be found analytically for any spherically symmetric metric by applying the above conditions to find
\begin{align}
\begin{split}
&2r_{ps} A(r_{ps})-A'(r)\vert_{r=r_{ps}} =0 ,
\\
& \hspace{0.5cm} b_{c} =\frac{r_{ps}}{\sqrt{A_{ps}}}
\end{split}
\end{align}
where $A_{ps} \equiv A(r_{ps})$. The first condition defines implicitly the photon sphere, while the second finds the associated (critical) impact parameter when evaluated there.

Light rays whose impact parameters satisfy $b \gtrsim b_c$ suffer heavy gravitational deflections when approaching the photon sphere such that they create, on the observer's screen, a series of self-similar {\it photon rings} indexed by the number of half-turns $n$ performed around the black hole, and located within lensing bands \cite{Cardenas-Avendano:2023obg} (i.e. photons that have turned at least $n$ times). Such photon rings approach, in the limit $n \to \infty$, a theoretical critical curve, corresponding to the projection, on the observer's plate image, of the photon sphere. The central brightness depression  caused by the much shorter path-length of those photons hitting the black hole horizon (i.e. those with $b<b_c$) might be naively associated with the critical curve (very frequently in the literature the inner region to such a curve is actually dubbed as the shadow), but this happens to be so only in spherical accretion disk models \cite{Vincent:2022fwj}, which is far from the more realistic situation of thick-accretion disks.

\subsection{Thin-disk modeling}

We consider an optically and geometrically thin accretion disk emitting monotonically in its own frame with a given (fixed) frequency $v_e$ and intensity $I_{e}=I(r) \delta(\nu - \nu_e)$\footnote{The setting employed here has been studied in several works; for the sake of this paper we refer the reader to the discussion of \cite{Olmo:2023lil}.}. The first assumption means that the disk is considered  transparent to its own radiation to some extent (for our purposes, we consider photons that cannot be absorbed by the disk before performing up to two half-turns, $n=2$), while the second assumption means that the thickness of the disk is negligible as compared to its spatial extension. In this modeling, using the conservation of the flux $I_{\nu_o}/\nu_o=I_{\nu_e}/\nu_e$,  and bearing in mind that the observer's frequency, $\nu_o$ is related to the emitted one via $\nu_o=g \nu_e$, with $g =A^{1/2}(r)$, then one can compute the total observed intensity via the formula
\begin{eqnarray}
    I(r)&=&\int d\nu_o I_o= \int d\nu_o g^3 I_{e}= \int d\nu_e g^4 I_e \nonumber \\
    &=&\sum_{n=0}^{2} A^2(r) I(r) \Big \vert_{r=r_n(b)}
\end{eqnarray}
where in the last expression we have used the monochromatic character of the emission and the fact that quantities are evaluated according to the transfer function $r=r_n(b)$. The latter correlates the radial location $r_n(b)$ on the disk that a photon with impact parameter $b$ has in its $n$-th half-turn around the black hole. Because we are assuming trajectories up to $n=2$ (as indicated by the sum symbol), the transfer function will be split into three different contributions: $n=0$ (hereafter referred to as the disk's {\it direct} emission), representing the first interaction of the photon with the disk from which it emanates; $n=1$ represents the lensed image from the back of the disk; and $n=2$ corresponds to the photon rings, according to the number of half-turns performed when hitting the disk.

\begin{figure}[t!]
\includegraphics[width=0.45\textwidth]{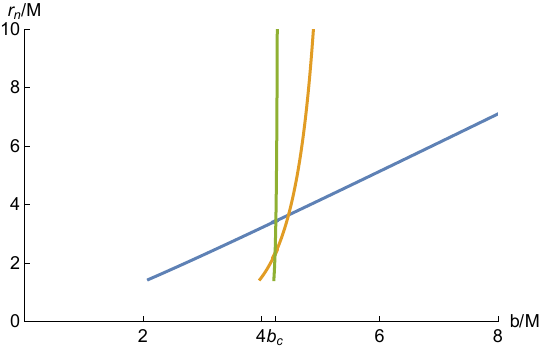}
\caption{The transfer function $r_n(b)$ for the direct ($n=0$, blue), and the first ($n=1$, orange) and second ($n=2$, green) photon ring emissions.}
\label{fig:trafuc}
\end{figure}

In Fig. \ref{fig:trafuc} we depict the transfer function $r_n(b)$ as a function of $b/M$ for scale-invariant black holes with $\epsilon=1.0$. For this value of the parameter $\epsilon$ one finds that
\begin{equation}
\frac{b_c^{\epsilon=1.0}}{M} \approx 4.23225
\end{equation} 
which is a significant reduction as compared to the Schwarzschild value, $b_c^S=3\sqrt{3} \approx 5.196$, while the photon sphere radius is reduced down from the Schwarzschild value $r_{ps}^S=3M$ to 
\begin{equation}
\frac{r_{ps}^{\epsilon=1.0}}{M} \approx 2.25574
\end{equation} Interestingly, given the fact that the critical impact parameter coincides with the shadow's radius employed in the methodology of the EHT to constrain it via the obtained data (via a correlation with the angular size of the ring of radiation enclosing it) of the M87 and Sgr A$^*$ images (equation such a radius with the critical impact parameter, in suitable units), it turns out that scale-invariant black holes with this choice of $\epsilon=1.0$ falls within the EHT constraint $4.21 \lesssim r_{sh}/M \lesssim 5.56$ \cite{EventHorizonTelescope:2022xqj}. The transfer function is not only needed in the process of generation of images (by linking impact parameter and radial distance) but also provides relevant information (via its slope) regarding the degree of demagnification of successive images of the accretion flow.

\subsection{Generation of images}

To generate images, we consider the pool of models introduced as suitable adaptations of Johnson's Standard Unbound function, and aimed to reproduce specific scenarios of the accretion flow in a simplified setting. Such models are given by the function \cite{Paugnat:2022qzy}
\begin{equation}
    I(r)=\frac{\exp\left(-\frac{1}{2}(\gamma + \text{arcosinh}(\left(\frac{r-\mu}{\sigma}^2 \right) \right)}{\sqrt{(r-\mu)^2 + \sigma^2}}
\end{equation}
and are characterized by three parameters corresponding to different properties of the emission profile: $\mu$ determines the location of its peak, \textcolor{purple}{$\sigma$} its width, and $\gamma$ its asymmetry. We consider an infinitely-thin disk located (without any loss of generality) in the equatorial plane $\theta=\pi/2$ and with a face-on orientation between its axis and the observer's line of sight. Among the pool of models previously considered in the literature, we focus on those studied by Gralla-Lupsasca-Marrone (GLM) introduced in \cite{Gralla:2020srx} and, in particular, those which peak at the innermost stable circular orbit for time-like particles (GLM3 model of \cite{Gralla:2020srx}). This is so because, though the disk is expected to extend up to the event horizon (GLM1/GLM2 models of \cite{Gralla:2020srx}), only when the disk's intensity peak is located outside enough of the photon sphere will the photon rings appear clearly isolated from the direct emission. This will allow us to study the modifications to the structure of such photon rings in a cleaner scenario. 

\begin{figure*}[t!]	
\includegraphics[width=6cm,height=4cm]{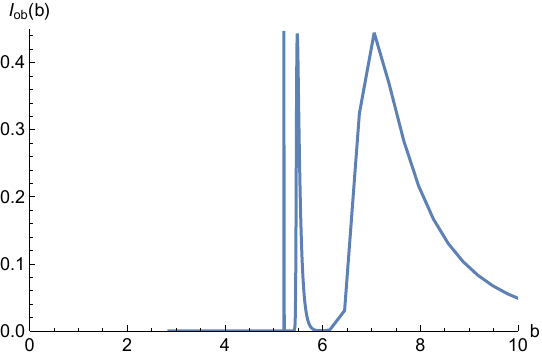}
\includegraphics[width=6cm,height=4cm]{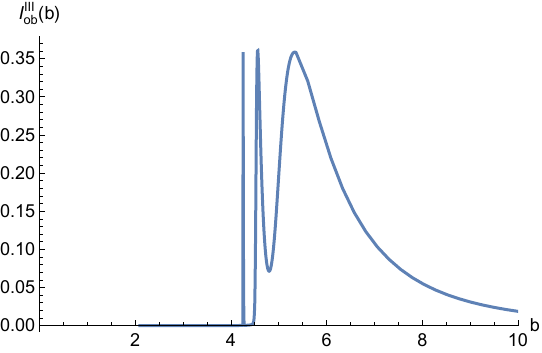}\\
\includegraphics[width=5.5cm,height=4cm]{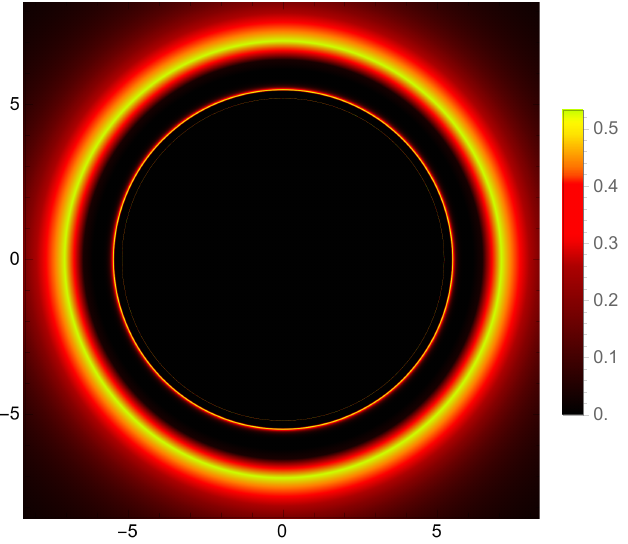}
\includegraphics[width=5.5cm,height=4cm]{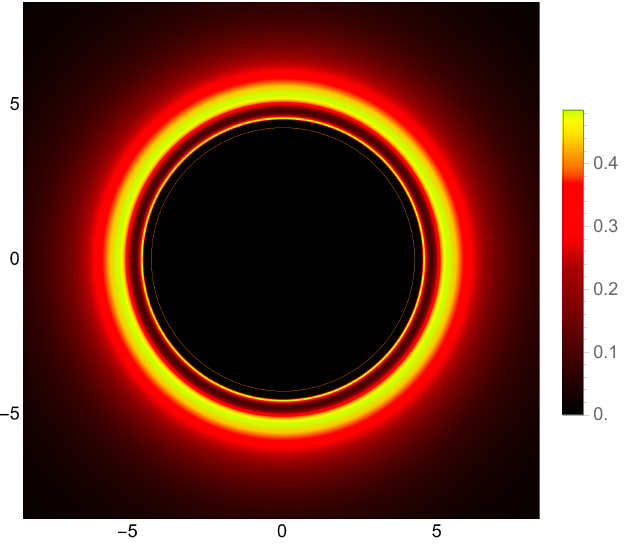} \\
\includegraphics[width=5.5cm,height=4cm]{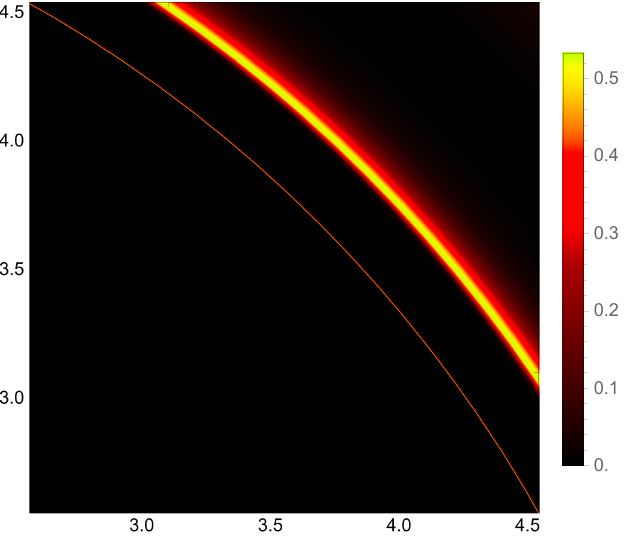}
\includegraphics[width=5.5cm,height=4cm]{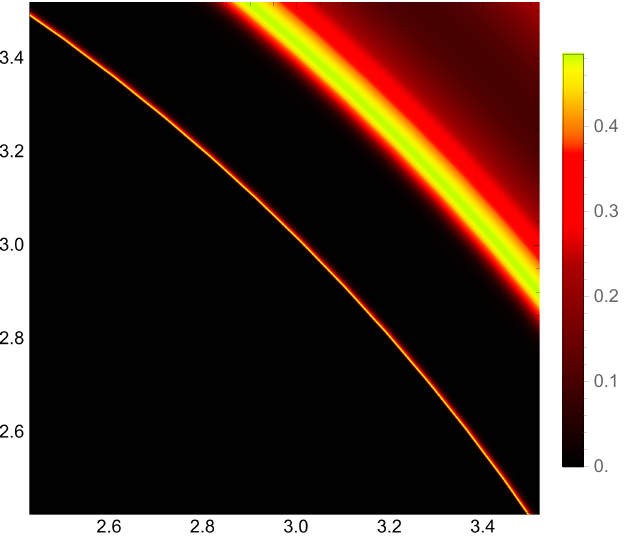}
\caption{The observed intensity (upper figures), the optical appearance (middle figures) and a cut of the first and second photon rings (bottom figures) for the Schwarzschild solution (left) and the scale-invariant black hole with $\epsilon=1.0$  (right).}
\label{fig:optical1}
\end{figure*}

In Fig. \ref{fig:optical1} we depict the optical appearance and the observed intensity of a scale-invariant black hole with $\epsilon=1.0$ (right figure) as compared to a Schwarzschild solution (left) figure. There are notable differences not only at the level of the reduction of the size of the central brightness depression (as already expected) but also significant changes in the location of the photon rings as compared to the direct emission as well as to their relative luminosities. Indeed, a glance at the corresponding observed intensities tells us that photon rings are much closer to the direct emission in the scale-invariant black hole than in the Schwarzschild one, and furthermore they seem to be more concentrated in $b$ (i.e. thinner). These qualitative differences can also be transferred into quantitative ones. 

On purely theoretical grounds, successive photon rings obey various relations governed by a series of {\it critical exponents} controlling their relative diameters, widths, or time elapsed between images (see the detailed analysis of \cite{Kocherlakota:2024hyq}). Of particular interest (for the setting explored here) is the so-called {\it Lyapunov exponent}, which is associated to the instability scale of nearly-bound geodesics, i.e., those with $r_o=r_{ps} + \delta r$, with $\delta r \ll r_{ps}$. This exponent has been computed elsewhere (see e.g. \cite{Macedo:2024qky}) and for the line element (\ref{m1}) reads as
\begin{equation}
    \gamma_{ps}= \pi \frac{1}{B_{ps}^{1/2} A'_{ps}}  \left[A_{ps}'^2-2A_{ps}A_{ps}''\right]^{1/2} \ .
\end{equation}
This quantity governs the exponential drift away of successive half-orbits, i.e. $r=r_o e^{\gamma_{ps} \phi/(2\pi)}$. It is important to note that this exponent is determined by the background geometry alone, thus being independent on the physics of the disk. Furthermore, successive photon rings obey an exponential-like relation of the form (in face-on orientation \cite{Kocherlakota:2024hyq})
\begin{equation} \label{eq:intlya}
\frac{I_{n+1}}{I_n} \approx e^{-\gamma_{ps}}   
\end{equation}
between the total flux intensity on the observer's screen of the $(n+1)$ photon ring as compared to that of the $n$ one. For instance, in the Schwarzschild case this number reads as $\gamma_{ps} =\pi$ while in the scale-invariant black hole case with $\epsilon=1.0$ it is lowered down to
\begin{equation}
\gamma_{ps} \approx 2.67
\end{equation}
implying, due to the exponential relation (\ref{eq:intlya}) a expectation of a factor $\sim 1.5$ of higher relative brightness in the scale-invariant black hole case than in the Schwarzschild one. This is in agreement with past research in the literature in which reductions in the shadow's size typically come alongside with relatively brighter photon rings.

The Lyapunov exponent is particularly relevant when framed with the possibility opened, by future upgrades of very-long baseline interferometry, like the ngEHT one \cite{Ayzenberg:2023hfw} or the Black Hole Explorer project \cite{Galison:2024bop}, of precise measurements of the features of the $n=1$ and $n=2$ photon rings. Though the relation (\ref{eq:intlya}) is only valid in the limit $n \to \infty$, research shows that in specific black hole cases (like the one considered here) the difference between the Lyapunov exponent and the approximation $I_2/I_1$ typically lies within a mere $\lesssim 1 \%$. Therefore, if such a measurement were to become a reality technologically at some point, then one would have a unique opportunity to read off directly the Lyapunov exponent from the photon rings features and put it back in correspondence with the predictions of the GR solutions or with any of its alternatives. In doing so, each method of measuring the width/diameter/luminosity has its disadvantages. 

Regarding luminosity of the successive photon rings, the universal expectation (\ref{eq:intlya}) would be true for a fully homogeneous flow in the region traveled by the $n=1$ and $n=2$ photon rings, which is not the actual case. In the simulations of (\ref{fig:optical1}) the actual intensity of each ring can be determined given the total ``area" spanned by each photon around its peak, and this amounts to an actual decrease of luminosity of $I_2/I_1 \approx 18.28$, which is significantly different from the expectation based on the Lyapunov exponent, $ \propto e^{\gamma_{ps}} \approx 14.77$. A similar situation happens in the Schwarzschild case, for which $e^{\gamma_{ps}} \approx 23.14$, but the ratio $I_1/I_2 \approx 27.83$. We can see this effect in the cut of the $n=1$ and $n=2$ photon rings of Fig  \ref{fig:optical1} (bottom figures). There we neatly see relatively more luminous and wider rings for the scale-invariant black hole (right figure) than in the Schwarzschild one (left figure). Similarly, we see how the innermost ring in the scale-invariant black hole, whose interior region defines the actual brightness depression, extends far beyond than on its Schwarzschild counterpart, and in both cases well inside the critical curve. This is consistent with studies showing that the actual shadow of a black hole with thin accretion disk can be reduced down, with a minimum absolute value given by the red-shifted image of the event horizon, see \cite{Chael:2021rjo}.

Furthermore, the above differences between theoretical expectations and actual luminosity ratios are typically far reduced for models in which the accretion flow takes its maximum value near the event horizon (see e.g. the analysis of \cite{daSilva:2023jxa}). However, in such a case photon rings tend to be overlapped with the direct emission rather than isolated from it (like in the model considered here), which risks spoiling the chances to neatly distinct photon rings between images.

\section{Quasi-normal modes-Shadows correspondence} \label{S:V}

One can realize at this stage that the potential driving black hole imaging, as given by Eq.(\ref{eq:V(r)s}), bears a close resemblance with the potential of QNMs, as given by Eq.(\ref{poten}) in the eikonal limit, $\ell \gg n$. This resemblance between the potentials driving each equation can be anchored into a more direct correspondence between physical quantities on each side. This is done (in the above limit) via the relation \cite{Stefanov:2010xz,Jusufi:2019ltj,Jusufi:2020dhz,Yang:2021zqy,Chen:2022nlw,Pedrotti:2024znu} (an in-depth, critical evaluation of this correspondence can be found at \cite{Pedrotti:2025upg})
\begin{equation} \label{eq:corrfull}
    \omega_{QNM} = \omega_R +i \omega_I = \Omega_c \left(\ell+\frac{1}{2} \right) -i \left(n + \frac{1}{2} \right) \vert \lambda \vert
\end{equation}
In this expression there are two quantities of interest related to black hole imaging. On the one hand, 
\begin{equation}
\Omega_c \equiv \frac{\dot{\varphi}}{\dot{t}}= \frac{A_{ps}^{1/2}}{r_{ps}}  
\end{equation}
is the angular velocity of unstable bound null geodesics, and can be shown to be  simply the inverse critical impact parameter, that is
\begin{equation} \label{eq:corr1}
\Omega_c=\frac{1}{b_c}
\end{equation}
As for $\lambda$, it is dubbed as the {\it principal Lyapunov exponent} since it provides a typical orbit time-scale $\propto $1/$\lambda$ of successive photon rings \cite{Cardoso:2008bp}. In fact, a little algebra involving several scaling relations allows to show that $\gamma_{ps}$ and $\lambda$ are related as
\begin{equation} \label{eq:corr2}
\vert \lambda \vert= \frac{\gamma_{ps}}{\pi b_c}
\end{equation}

The correspondence is further closed by recognizing that the shadow's radius, interpreted as the region bounded by the critical curve in the observer's screen, is found as
\begin{equation} \label{eq:omegar}
r_{sh} \overset{\ell  \gg n}{=} \frac{\left(\ell +\frac{1}{2}\right)}{\omega_R} 
\end{equation}
while we point out that this is not the actual brightness depression found in images, which is model-dependent.

As for $\lambda$ and $\gamma_{ps}$, both are related to unstable bound geodesics but while the form captures the coordinate time elapsed in the exponential drift of successive orbits, the latter does it in terms of the number of half-turns $n$, thus providing a more direct and suitable link to photon rings and QNMs via the relation
\begin{equation} \label{eq:omegai}
\gamma_{ps}=\frac{\pi b_c  \omega_I}{\left(n+\frac{1}{2} \right)}
\end{equation}

These are remarkable relations linking two seemingly different phenomena: QNM frequencies as resulting from the perturbation upon a space-time metric, and photon rings/shadows as the  imaging of light rays, which nonetheless are weaved together in the geometric optics approximation provided by the eikonal limit. Indeed, while the real part of the QNM, providing its amplitude, correlates with the shadow's radius, the imaginary part, providing the damping time, does it so with the width/diameter/luminosity of photon rings. This, therefore, links observable quantities on each side of the correspondence. 

For the sake of this work we can check the accurateness of this correspondence with the examples at hand. To this end, we take several examples for QNMs frequency appearing in Table \ref{tab:1} and extract both $\Omega_c$ and $\vert \lambda \vert$ following the correspondence (\ref{eq:corrfull}) for both Schwarzschild ($\epsilon =0$) and scale-invariance black holes with $\epsilon=1.0$ taking $\ell=10$ and $n=0,1,2,3$. We subsequently extract from these quantities both the critical impact parameter inferred by QNMs, $b_c^{QNM}$, using Eq.(\ref{eq:corr1}), and the associated Lyapunov exponent, $\gamma_{ps}^{QNM}$, using Eq.(\ref{eq:omegai}), and compare these estimations with those values from direct black hole imaging, $b_c^{BHI}$ and $\gamma_{ps}^{BHI}$, as found before in this section. The results are depicted in Table \ref{tab:1}. As it can be seen there, the prediction from QNMs for these relevant quantities accurately reproduce those of direct black hole imaging with ratios of $b_{c}^{BHI}/b_c^{QNM}$ and $\gamma_{ps}^{BHI}/\gamma_{ps}^{QNM}$ very close to one, in particular, for the mode $n=0$ as expected from the limit of validity $\ell \gg n$ of the correspondence.

It should be stressed that this QNM-shadows correspondence is subjected to some limitations, essentially related to the WKB procedure itself, see \cite{Konoplya:2019hlu} (and also \cite{Pedrotti:2025upg}) for a discussion of this. In the present case, the agreement between each side decreases as higher overtone numbers are considered, something that was already expected.

\begin{table}[t]
\label{table:GWSH}
\begin{center}
\begin{tabular}{|c|c|c|c|c|c|c|} 
\hline
\textbf{Case} & \textbf{Overtone} & $\boldsymbol{\Omega_c}$ & $\boldsymbol{\vert \lambda \vert}$ & 
$\boldsymbol{b_{c}^{QNM}}$ &
$\boldsymbol{\frac{b_{c}^{BHI}}{b_c^{QNM}}}$ & $\boldsymbol{\frac{\gamma_{ps}^{BHI}}{\gamma_{ps}^{QNM}}}$ \\
\hline
\multirow{4}{*}{\makecell{$\epsilon = 0$\\$\ell = 10$}} & $n=0$ & 0.1925 & 0.1925 & 5.1946 & 1.0003 & 0.9999 \\
\cline{2-7}
                         & $n=1$ & 0.1920  & 0.1927  & 5.2077  & 0.9977  & 
                         0.9963 \\
\cline{2-7}
                         & $n=2$ & 
                         0.1910  &
                         0.1931  &
                         5.2337  &
                         0.9928  &
                         0.9892 \\
\cline{2-7}
                         & $n=3$ & 0.1896 & 0.1937 & 5.2728 & 0.9854 & 0.9787 \\
\hline
\multirow{4}{*}{\makecell{$\epsilon = 1.0$\\$\ell = 10$}} &  $n=0$     & 0.2363 &  0.2012 & 4.2309  & 0.9997 & 1.0070 \\
\cline{2-7}
                         &   $n=1$  & 0.2352 & 
                         0.2013 &
                         4.2373 &
                         0.9988 &
                         0.9978 \\
\cline{2-7}
                         &   $n=2$     &
                         0.2352 &
                         0.2016 &
                         4.2499 &
                         0.9958 &
                         0.9933 \\
\cline{2-7}
                         &   $n=3$     & 0.2342  & 0.2021 & 4.2688 & 0.9914 & 0.9868 \\
\hline
\end{tabular}
\caption{Values of the parameter $\Omega_c$ and of $\vert \lambda \vert$ found from the correspondence (\ref{eq:corrfull}), as well as the inferred values from QNM analysis on $b_c^{QNM}$ using (\ref{eq:corr1}) and the ratios $b_c^{BHI}/b_c^{QNM}$ and $\gamma_{ps}^{BHI}/\gamma_{ps}^{QNM}$ using (\ref{eq:omegai}). The values for the QNM frequencies are extracted from the scalar field perturbations of Table \ref{tab:1}.}
\end{center}
\end{table}


\section{Discussion and Conclusion}\label{S:VI}

In this work we have analyzed the QNMs and shadows of a scale-dependent regular black hole in four space-time dimensions. Such a model can be framed (within GR) with non-linear electrodynamics and scalar fields as the matter sources supporting it. 

Regarding QNMs, our study focused on two types of perturbations: (i) neutral scalar and (ii) Dirac fields. Since the black hole solution considered here is a slight deformation of the Schwarzschild geometry and leads to an effective potential of relatively simple structure, we used the semi-analytical WKB approximation to compute the associated QNM  frequencies. We confirmed that the effective potentials for both scalar and Dirac perturbations exhibit a barrier-type profile, allowing the application of the sixth-order WKB method to extract the QNM spectra. Our results are presented in two tables (one for scalar and one for Dirac perturbations) at Appendix \ref{A:QNM} and illustrated graphically in several figures, showing the dependence of the QNM frequencies on the scale-dependent parameter $\epsilon$, the angular momentum number $\ell/\xi$ and the overtone number $n$. Since the imaginary part of the frequencies remains negative in all cases, we conclude that the black hole is stable under both scalar and Dirac perturbations. To further strengthen our findings, we have derived analytical expressions for the QNM frequencies beyond the eikonal limit for both scalar and Dirac perturbations. The resulting modes show excellent agreement with the numerical values obtained using the WKB approximation.

It is worth stressing that both the real and the imaginary part have very well-defined meaning in term of physics of waves. 
Indeed, the real part of a QNM  frequency is related to the speed at which a disturbance oscillates, while the imaginary component is associated with the rate at which it dissipates.
In other words, these QNMs act as fingerprints of space-time, revealing information about its shape and stability, and provide insights into certain key aspects of black holes, such as their mass and spin, as well as offering clues about physics beyond the scope of general relativity. According to our numerical study, we observed that quantum corrections considerably alter the real part of the QNM  frequency, meaning that the ``quantum" effect is significant and the oscillation period is altered. Conversely, the imaginary part remains largely unaffected, and it is difficult to draw strong conclusions based on the observed behaviour.

Regarding black hole imaging/shadows, we ray-traced our scale-invariant metric for a reference value of $\epsilon=1.0$, chosen on the grounds of satisfying the recent constraints set by the EHT Collaboration on the calibrated size of the shadow region observed in the M87 and Sgr A$^*$ central objects. Using a thin-disk model with a monochromatic emission in the disk's frame and peaking near the innermost stable time-like orbit in order to neatly see the photon rings isolated from the disk's direct emission (instead of overlapped with it), we generated the image of such a black hole and compared it with the corresponding one of a Schwarzschild black hole. We found significant differences, besides the expected reduced size of the central brightness depression, in terms of less tightly packed photon rings, which further penetrate the critical curve region, and which are relatively more luminous. The latter was estimated according to the Lyapunov exponent of nearly-bound unstable geodesics, one of the exponents characterizing the actual features of successive photon rings, and later we double-checked it with the actual ratio of luminosity between the $n=1$ and $n=2$ photon rings, finding moderate differences with respect to the theoretical expectation due to the disk model peaking-region.

It should be noted that the Lyapunov exponent is one of the theoretical quantities characterizing unstable bound orbits \cite{Kocherlakota:2024hyq} and providing the sets of relevant observables (such as diameter, width, luminosity, time elapsed) of photon rings. The latter are the target of  future very-long baseline interferometric projects, like the next-generation EHT, or the Black Hole Explorer project \cite{Lupsasca:2024xhq}. 

Furthermore, both the angular velocity and the Lyapunov exponent of unstable bound geodesics play a key role within an intriguing correspondence between QNM frequencies (in the eikonal limit) and black hole imaging quantities (in the strong-deflection limit). Indeed, assuming such a correspondence, we were able to work out the latter quantities starting from the former quantities alone, to a great degree of accuracy (for modes $\ell \gg n$). This provides a specific example showing the latent power hidden in this correspondence, bypassing the question of whether this correspondence could work for theories beyond GR \cite{Chen:2022nlw}, though this is an issue to be settled on a case-by-case basis.

To conclude, while scale-invariant black holes may be seen as just a mere theoretical construction to deliver regular black holes, in this work we have used it to highlight the available opportunities from studying QNMs and light imaging for the same black hole candidate. While current technology at hand splits such measurements into two very distinct regimes, namely, dozens of solar masses for QNMs and at least $\sim 4.1 \times 10^6$ solar mass for shadows (the one of Sgr A$^*$), a black hole {\it multimessenger astronomy} seems a promising future path of research for theoreticians and experimentalists alike. 

In other words, although current gravitational-wave observations primarily involve stellar-mass black holes, future detectors such as LISA \cite{Barausse:2020rsu} will probe supermassive black holes in the same mass range as those observed by the Event Horizon Telescope. In such a sense, the proposed QNM-shadow correspondence could then be  a genuine multi-messenger test of black hole spacetimes.

By cross-testing both fields, this could allow to sort out current theoretical and observational uncertainties that plague (though in different ways) each messenger, requiring further work on the possibilities and limits of such a correspondence. We hope to further report on this issue soon.

\section*{Acknowledgments}

A.~R. would like to express his gratitude to Silesian University in Opava, Czech Republic, for their financial support.
A.~R. is very grateful for the hospitality of the University of Valencia (Spain) and the Complutense University of Madrid (Spain). 
The creation of this article was supported by the grant program Vouchers for Universities in the Moravian-Silesian Region (registration number CZ.10.03.01/00/23\_042/0000390).
This work is supported by the Spanish National
Grants PID2020-116567GB-C21, PID2022-138607NBI00, PID2023-149560NB-C21, and CNS2024-154444, funded by MICIU/AEI/10.13039/501100011033 (``ERDF A way of making Europe" and ``PGC Generaci\'on de Conocimiento").
We also acknoledge support from the Severo Ochoa Excellence Grant CEX2023-001292-S and the European Horizon Europe staff exchange (SE) programme HORIZON-MSCA-2021-SE-01 Grant No. NewFunFiCO-101086251. The authors would like to acknowledge the contribution of the COST Action CA23130 (``Bridging high and low energies in search of quantum gravity (BridgeQG)").

\appendix

\section{Computation of quasi-normal modes} \label{A:QNM}

In this Appendix we display the results for the computation of massless scalar QNMs in Table \ref{tab:1}, and massless Dirac QNMs in Table \ref{tab:2}. For each field we consider values of the scale-dependent gravity parameter $\epsilon$ in the interval $\epsilon \in [0,1.0]$ and for each overtone number $n=0,1,2,3$ we compute the real and imaginary parts of the QNM for values of the angular degree number $\ell=6,7,8,9,10$, 
for the scalar case, and $\xi=6,7,8,9,10$, for the Dirac case. Furthermore, we show the QNM frequencies using the exact analytic expressions in Table \ref{tab:3}, for the scalar case, and Table \ref{tab:4}, for the Dirac case, taking $M=1$.

\begin{table*}
    \centering
	\caption{Massless scalar quasinormal modes using the sixth-order WKB approximation.}
	\label{tab:1}       
	\begin{tabular}{c|c|c|c|c|c|c}
		\hline\noalign{\smallskip}
		$\epsilon$  & $n$ & $\ell=6$ & $\ell=7$ & $\ell=8$ & $\ell=9$ & $\ell=10$  \\
		\noalign{\smallskip}\hline\noalign{\smallskip}
          \noalign{\smallskip}\hline
                \rowcolor{purple!30}
        0.0 & 0 & 1.25189\, -0.0963051 i & 1.44421\, -0.0962852 i & 1.63656\, -0.0962719 i & 1.82893\, -0.0962626 i & 2.02132\, -0.0962558 i \\
        0.0 & 1 & 1.24375\, -0.2897360 i & 1.43714\, -0.2894730 i & 1.63031\, -0.2892970 i & 1.82333\, -0.2891730 i & 2.01625\, -0.2890830 i \\
        0.0 & 2 & 1.22784\, -0.4856010 i & 1.42323\, -0.4844980 i & 1.61797\, -0.4837570 i & 1.81225\, -0.4832350 i & 2.00620\, -0.4828540 i \\
        0.0 & 3 & 1.20489\, -0.6854090 i & 1.40297\, -0.6825240 i & 1.59987\, -0.6805720 i & 1.79592\, -0.6791920 i & 1.99133\, -0.6781830 i \\
		\noalign{\smallskip}\hline    
                \rowcolor{purple!30}
        0.1 & 0 & 1.27330\, -0.0968364 i & 1.46891\, -0.0968168 i & 1.66455\, -0.0968038 i & 1.86021\, -0.0967946 i & 2.05589\, -0.0967879 i \\
        0.1 & 1 & 1.26533\, -0.2913130 i & 1.46198\, -0.2910560 i & 1.65842\, -0.2908830 i & 1.85472\, -0.2907620 i & 2.05092\, -0.2906730 i \\
        0.1 & 2 & 1.24973\, -0.4881730 i & 1.44835\, -0.4870930 i & 1.64633\, -0.4863670 i & 1.84386\, -0.4858560 i & 2.04106\, -0.4854820 i \\
        0.1 & 3 & 1.22723\, -0.6888900 i & 1.42848\, -0.6860670 i & 1.62858\, -0.6841560 i & 1.82784\, -0.6828060 i & 2.02648\, -0.6818170 i \\
		\noalign{\smallskip}\hline
                \rowcolor{purple!30}
        0.2 & 0 & 1.29587\, -0.0973619 i & 1.49494\, -0.0973427 i & 1.69404\, -0.0973299 i & 1.89316\, -0.0973209 i & 2.09230\, -0.0973143 i \\
        0.2 & 1 & 1.28806\, -0.2928720 i & 1.48815\, -0.2926200 i & 1.68804\, -0.2924510 i & 1.88779\, -0.2923320 i & 2.08744\, -0.2922460 i \\
        0.2 & 2 & 1.27279\, -0.4907110 i & 1.47481\, -0.4896550 i & 1.67620\, -0.4889450 i & 1.87715\, -0.4884450 i & 2.07779\, -0.4880800 i \\
        0.2 & 3 & 1.25077\, -0.6923160 i & 1.45536\, -0.6895580 i & 1.65883\, -0.6876920 i & 1.86148\, -0.6863720 i & 2.06352\, -0.6854060 i \\
		\noalign{\smallskip}\hline
                \rowcolor{purple!30}
        0.3 & 0 & 1.31969\, -0.0978783 i & 1.52242\, -0.0978595 i & 1.72518\, -0.0978469 i & 1.92796\, -0.0978381 i & 2.13076\, -0.0978317 i \\
        0.3 & 1 & 1.31206\, -0.2944020 i & 1.51579\, -0.2941560 i & 1.71932\, -0.2939910 i & 1.92271\, -0.2938750 i & 2.12600\, -0.2937900 i \\
        0.3 & 2 & 1.29715\, -0.4931960 i & 1.50275\, -0.4921670 i & 1.70775\, -0.4914740 i & 1.91232\, -0.4909870 i & 2.11657\, -0.4906310 i \\
        0.3 & 3 & 1.27563\, -0.6956610 i & 1.48375\, -0.6929730 i & 1.69078\, -0.6911530 i & 1.89700\, -0.6898670 i & 2.10263\, -0.6889250 i \\
		\noalign{\smallskip}\hline
                \rowcolor{purple!30}
        0.4 & 0 & 1.34492\, -0.0983812 i & 1.55151\, -0.0983629 i & 1.75815\, -0.0983506 i & 1.96480\, -0.0983420 i & 2.17147\, -0.0983357 i \\
        0.4 & 1 & 1.33748\, -0.2958900 i & 1.54505\, -0.2956500 i & 1.75243\, -0.2954900 i & 1.95968\, -0.2953770 i & 2.16684\, -0.2952950 i \\
        0.4 & 2 & 1.32294\, -0.4956080 i & 1.53234\, -0.4946060 i & 1.74115\, -0.4939320 i & 1.94955\, -0.4934580 i & 2.15764\, -0.4931110 i \\
        0.4 & 3 & 1.30196\, -0.6988930 i & 1.51381\, -0.6962800 i & 1.72461\, -0.6945100 i & 1.93462\, -0.6932590 i & 2.14405\, -0.6923420 i \\
		\noalign{\smallskip}\hline
                \rowcolor{purple!30}
        0.5 & 0 & 1.37170\, -0.0988651 i & 1.58241\, -0.0988472 i & 1.79315\, -0.0988352 i & 2.00392\, -0.0988269 i & 2.21470\, -0.0988207 i \\
        0.5 & 1 & 1.36446\, -0.2973200 i & 1.57612\, -0.2970870 i & 1.78759\, -0.2969310 i & 1.99894\, -0.2968210 i & 2.21019\, -0.2967410 i \\
        0.5 & 2 & 1.35032\, -0.4979170 i & 1.56375\, -0.4969450 i & 1.77662\, -0.4962910 i & 1.98908\, -0.4958300 i & 2.20125\, -0.4954940 i \\
        0.5 & 3 & 1.32991\, -0.7019710 i & 1.54574\, -0.6994370 i & 1.76053\, -0.6977210 i & 1.97456\, -0.6965070 i & 2.18803\, -0.6956180 i \\
		\noalign{\smallskip}\hline
        0.6 & 0 & 1.40022\, -0.0993227 i & 1.61531\, -0.0993053 i & 1.83044\, -0.0992936 i & 2.04559\, -0.0992855 i & 2.26075\, -0.0992795 i \\
        0.6 & 1 & 1.39320\, -0.2986690 i & 1.60921\, -0.2984430 i & 1.82505\, -0.2982920 i & 2.04076\, -0.2981860 i & 2.25638\, -0.2981080 i \\
        0.6 & 2 & 1.37948\, -0.5000850 i & 1.59722\, -0.4991450 i & 1.81440\, -0.4985130 i & 2.03120\, -0.4980670 i & 2.24771\, -0.4977420 i \\
        0.6 & 3 & 1.35969\, -0.7048410 i & 1.57974\, -0.7023930 i & 1.79879\, -0.7007340 i & 2.01711\, -0.6995610 i & 2.23488\, -0.6987010 i \\
		\noalign{\smallskip}\hline
        0.7 & 0 & 1.43072\, -0.0997443 i & 1.65049\, -0.0997274 i & 1.87030\, -0.0997161 i & 2.09014\, -0.0997082 i & 2.30999\, -0.0997024 i \\
        0.7 & 1 & 1.42393\, -0.2999080 i & 1.64459\, -0.2996900 i & 1.86509\, -0.2995440 i & 2.08547\, -0.2994420 i & 2.30576\, -0.2993670 i \\
        0.7 & 2 & 1.41067\, -0.5020640 i & 1.63300\, -0.5011590 i & 1.85480\, -0.5005490 i & 2.07622\, -0.5001200 i & 2.29737\, -0.4998070 i \\
        0.7 & 3 & 1.39153\, -0.7074330 i & 1.61610\, -0.7050770 i & 1.83971\, -0.7034800 i & 2.06260\, -0.7023500 i & 2.28497\, -0.7015230 i \\
		\noalign{\smallskip}\hline
        0.8 & 0 & 1.46346\, -0.1001170 i & 1.68826\, -0.1001010 i & 1.91310\, -0.1000900 i & 2.13796\, -0.1000820 i & 2.36285\, -0.1000770 i \\
        0.8 & 1 & 1.45692\, -0.3009980 i & 1.68258\, -0.3007890 i & 1.90808\, -0.3006490 i & 2.13346\, -0.3005500 i & 2.35877\, -0.3004790 i \\
        0.8 & 2 & 1.44414\, -0.5037880 i & 1.67141\, -0.5029200 i & 1.89817\, -0.5023360 i & 2.12456\, -0.5019240 i & 2.35069\, -0.5016230 i \\
        0.8 & 3 & 1.42570\, -0.7096530 i & 1.65513\, -0.7073960 i & 1.88363\, -0.7058660 i & 2.11144\, -0.7047840 i & 2.33875\, -0.7039901 i \\
		\noalign{\smallskip}\hline
        0.9 & 0 & 1.49877\, -0.1004230 i & 1.72900\, -0.1004080 i & 1.95926\, -0.1003970 i & 2.18955\, -0.1003900 i & 2.41987\, -0.1003840 i \\
        0.9 & 1 & 1.49250\, -0.3018860 i & 1.72355\, -0.3016870 i & 1.95445\, -0.3015530 i & 2.18524\, -0.3014590 i & 2.41596\, -0.3013910 i \\
        0.9 & 2 & 1.48025\, -0.5051670 i & 1.71284\, -0.5043400 i & 1.94494\, -0.5037830 i & 2.17670\, -0.5033900 i & 2.40821\, -0.5031040 i \\
        0.9 & 3 & 1.46256\, -0.7113730 i & 1.69723\, -0.7092240 i & 1.93100\, -0.7077680 i & 2.16412\, -0.7067370 i & 2.39676\, -0.7059810 i \\
		\noalign{\smallskip}\hline
        1.0 & 0 & 1.53707\, -0.1006390 i & 1.77319\, -0.1006240 i & 2.00934\, -0.1006140 i & 2.24552\, -0.1006070 i & 2.48172\, -0.1006020 i \\
        1.0 & 1 & 1.53109\, -0.3025000 i & 1.76799\, -0.3023110 i & 2.00475\, -0.3021840 i & 2.24141\, -0.3020950 i & 2.47799\, -0.3020300 i \\
        1.0 & 2 & 1.51940\, -0.5060780 i & 1.75777\, -0.5052950 i & 1.99568\, -0.5047680 i & 2.23326\, -0.5043970 i & 2.47061\, -0.5041260 i \\
        1.0 & 3 & 1.50252\, -0.7124180 i & 1.74287\, -0.7103870 i & 1.98238\, -0.7090100 i & 2.22126\, -0.7080360 i & 2.45968\, -0.7073220 i \\
		\noalign{\smallskip}\hline
	\end{tabular}
\end{table*}


\begin{table*}
    \centering
	\caption{Massless Dirac quasinormal modes using the sixth-order WKB approximation.}
	\label{tab:2}       
	\begin{tabular}{c|c|c|c|c|c|c}
		\hline\noalign{\smallskip}
		$\epsilon$  & $n$ & $\xi=6$ & $\xi=7$ & $\xi=8$ & $\xi=9$ & $\xi=10$  \\
		\noalign{\smallskip}\hline\noalign{\smallskip}
          \noalign{\smallskip}\hline
                \rowcolor{olive!30}
        0.0 & 0 & 1.15307\, -0.0962450 i & 1.34575\, -0.0962397 i & 1.53838\, -0.0962363 i & 1.73096\, -0.0962339 i & 1.92352\, -0.0962322 i \\
        0.0 & 1 & 1.14425\, -0.2897000 i & 1.33817\, -0.2894290 i & 1.53173\, -0.2892530 i & 1.72505\, -0.2891320 i & 1.91819\, -0.2890450 i \\
        0.0 & 2 & 1.12707\, -0.4860100 i & 1.32330\, -0.4847270 i & 1.51864\, -0.4838890 i & 1.71336\, -0.4833120 i & 1.90764\, -0.4828980 i \\
        0.0 & 3 & 1.10246\, -0.6869170 i & 1.30172\, -0.6834570 i & 1.49948\, -0.6811790 i & 1.69616\, -0.6796020 i & 1.89206\, -0.6784680 i \\
        \noalign{\smallskip}\hline
                \rowcolor{olive!30}
        0.1 & 0 & 1.17283\, -0.0967773 i & 1.36880\, -0.0967721 i & 1.56471\, -0.0967687 i & 1.76059\, -0.0967664 i & 1.95643\, -0.0967648 i \\
        0.1 & 1 & 1.16418\, -0.2912770 i & 1.36137\, -0.2910120 i & 1.55820\, -0.2908390 i & 1.75479\, -0.2907200 i & 1.95121\, -0.2906360 i \\
        0.1 & 2 & 1.14734\, -0.4885710 i & 1.34679\, -0.4873160 i & 1.54536\, -0.4864950 i & 1.74333\, -0.4859310 i & 1.94086\, -0.4855260 i \\
        0.1 & 3 & 1.12322\, -0.6903630 i & 1.32564\, -0.6869790 i & 1.52658\, -0.6847490 i & 1.72647\, -0.6832070 i & 1.92559\, -0.6820960 i \\     \noalign{\smallskip}\hline
                \rowcolor{olive!30}
        0.2 & 0 & 1.19365\, -0.0973038 i & 1.39308\, -0.0972988 i & 1.59246\, -0.0972955 i & 1.79180\, -0.0972932 i & 1.99111\, -0.0972916 i \\
        0.2 & 1 & 1.18518\, -0.2928360 i & 1.38581\, -0.2925760 i & 1.58608\, -0.2924080 i & 1.78612\, -0.2922920 i & 1.98600\, -0.2922090 i \\
        0.2 & 2 & 1.16870\, -0.4910980 i & 1.37154\, -0.4898720 i & 1.57352\, -0.4890700 i & 1.77490\, -0.4885180 i & 1.97587\, -0.4881220 i \\
        0.2 & 3 & 1.14509\, -0.6937510 i & 1.35083\, -0.6904470 i & 1.55513\, -0.6882700 i & 1.75840\, -0.6867630 i & 1.96092\, -0.6856780 i \\  
		\noalign{\smallskip}\hline
                \rowcolor{olive!30}
        0.3 & 0 & 1.21563\, -0.0978214 i & 1.41873\, -0.0978164 i & 1.62176\, -0.0978132 i & 1.82476\, -0.0978110 i & 2.02773\, -0.0978095 i \\
        0.3 & 1 & 1.20736\, -0.2943660 i & 1.41162\, -0.2941130 i & 1.61553\, -0.2939480 i & 1.81921\, -0.2938350 i & 2.02274\, -0.2937540 i \\
        0.3 & 2 & 1.19126\, -0.4935730 i & 1.39768\, -0.4923770 i & 1.60325\, -0.4915950 i & 1.80825\, -0.4910570 i & 2.01284\, -0.4906710 i \\
        0.3 & 3 & 1.16819\, -0.6970570 i & 1.37745\, -0.6938370 i & 1.58529\, -0.6917160 i & 1.79213\, -0.6902470 i & 1.99823\, -0.6891890 i \\
		\noalign{\smallskip}\hline
                \rowcolor{olive!30}
        0.4 & 0 & 1.23891\, -0.0983255 i & 1.44588\, -0.0983207 i & 1.65279\, -0.0983176 i & 1.85966\, -0.0983155 i & 2.06650\, -0.0983139 i \\
        0.4 & 1 & 1.23085\, -0.2958540 i & 1.43895\, -0.2956080 i & 1.64671\, -0.2954480 i & 1.85425\, -0.2953380 i & 2.06163\, -0.2952590 i \\
        0.4 & 2 & 1.21515\, -0.4959720 i & 1.42535\, -0.4948090 i & 1.63474\, -0.4940490 i & 1.84356\, -0.4935250 i & 2.05198\, -0.4931500 i \\
        0.4 & 3 & 1.19266\, -0.7002470 i & 1.40563\, -0.6971180 i & 1.61723\, -0.6950550 i & 1.82784\, -0.6936270 i & 2.03774\, -0.6925980 i \\
        \noalign{\smallskip}\hline
                \rowcolor{olive!30}
        0.5 & 0 & 1.26363\, -0.0988107 i & 1.47471\, -0.0988060 i & 1.68573\, -0.0988030 i & 1.89671\, -0.0988009 i & 2.10767\, -0.0987995 i \\
        0.5 & 1 & 1.25579\, -0.2972840 i & 1.46797\, -0.2970450 i & 1.67982\, -0.2968890 i & 1.89145\, -0.2967830 i & 2.10293\, -0.2967060 i \\
        0.5 & 2 & 1.24052\, -0.4982680 i & 1.45474\, -0.4971400 i & 1.66817\, -0.4964030 i & 1.88106\, -0.4958950 i & 2.09355\, -0.4955300 i \\
        0.5 & 3 & 1.21864\, -0.7032800 i & 1.43556\, -0.7002480 i & 1.65115\, -0.6982480 i & 1.86577\, -0.6968630 i & 2.07969\, -0.6958660 i \\
	\noalign{\smallskip}\hline
        0.6 & 0 & 1.28997\, -0.0992696 i & 1.50542\, -0.0992651 i & 1.72082\, -0.0992622 i & 1.93619\, -0.0992602 i & 2.15153\, -0.0992588 i \\
        0.6 & 1 & 1.28236\, -0.2986320 i & 1.49888\, -0.2984010 i & 1.71509\, -0.2982510 i & 1.93109\, -0.2981480 i & 2.14693\, -0.2980740 i \\
        0.6 & 2 & 1.26755\, -0.5004220 i & 1.48606\, -0.4993330 i & 1.70380\, -0.4986200 i & 1.92100\, -0.4981290 i & 2.13783\, -0.4977760 i \\
        0.6 & 3 & 1.24633\, -0.7061010 i & 1.46745\, -0.7031730 i & 1.68728\, -0.7012420 i & 1.90618\, -0.6999030 i & 2.12439\, -0.6989390 i \\
		\noalign{\smallskip}\hline
        0.7 & 0 & 1.31812\, -0.0996927 i & 1.53826\, -0.0996884 i & 1.75834\, -0.0996855 i & 1.97839\, -0.0996836 i & 2.19842\, -0.0996822 i \\
        0.7 & 1 & 1.31077\, -0.2998710 i & 1.53194\, -0.2996490 i & 1.75280\, -0.2995040 i & 1.97346\, -0.2994050 i & 2.19398\, -0.2993330 i \\
        0.7 & 2 & 1.29644\, -0.5023860 i & 1.51954\, -0.5013370 i & 1.74188\, -0.5006510 i & 1.96371\, -0.5001780 i & 2.18518\, -0.4998390 i \\
        0.7 & 3 & 1.27593\, -0.7086410 i & 1.50155\, -0.7058250 i & 1.72591\, -0.7039660 i & 1.94938\, -0.7026780 i & 2.17218\, -0.7017500 i \\
		\noalign{\smallskip}\hline
        0.8 & 0 & 1.34835\, -0.1000670 i & 1.57351\, -0.1000630 i & 1.79863\, -0.1000600 i & 2.02371\, -0.1000580 i & 2.24877\, -0.1000570 i \\
        0.8 & 1 & 1.34127\, -0.3009610 i & 1.56743\, -0.3007480 i & 1.79329\, -0.3006090 i & 2.01896\, -0.3005140 i & 2.24449\, -0.3004460 i \\
        0.8 & 2 & 1.32747\, -0.5040920 i & 1.55548\, -0.5030880 i & 1.78277\, -0.5024310 i & 2.00957\, -0.5019780 i & 2.23601\, -0.5016530 i \\
        0.8 & 3 & 1.30770\, -0.7108040 i & 1.53815\, -0.7081080 i & 1.76739\, -0.7063290 i & 1.99576\, -0.7050960 i & 2.22349\, -0.7042070 i \\
       \noalign{\smallskip}\hline
        0.9 & 0 & 1.38096\, -0.1003750 i & 1.61155\, -0.1003710 i & 1.84209\, -0.1003690 i & 2.07260\, -0.1003670 i & 2.30308\, -0.1003660 i \\
        0.9 & 1 & 1.37417\, -0.3018490 i & 1.60571\, -0.3016460 i & 1.83697\, -0.3015140 i & 2.06804\, -0.3014230 i & 2.29898\, -0.3013580 i \\
        0.9 & 2 & 1.36094\, -0.5054520 i & 1.59426\, -0.5044970 i & 1.82689\, -0.5038710 i & 2.05904\, -0.5034400 i & 2.29085\, -0.5031310 i \\
        0.9 & 3 & 1.34197\, -0.7124620 i & 1.57764\, -0.7098970 i & 1.81213\, -0.7082040 i & 2.04579\, -0.7070310 i & 2.27884\, -0.7061850 i \\
		\noalign{\smallskip}\hline
        1.0 & 0 & 1.41634\, -0.1005920 i & 1.65281\, -0.1005890 i & 1.88924\, -0.1005860 i & 2.12563\, -0.1005850 i & 2.36200\, -0.1005840 i \\
        1.0 & 1 & 1.40986\, -0.3024610 i & 1.64724\, -0.3022700 i & 1.88436\, -0.3021450 i & 2.12129\, -0.3020590 i & 2.35809\, -0.3019980 i \\
        1.0 & 2 & 1.39724\, -0.5063420 i & 1.63632\, -0.5054400 i & 1.87473\, -0.5048490 i & 2.11270\, -0.5044420 i & 2.35033\, -0.5041490 i \\
        1.0 & 3 & 1.37913\, -0.7134380 i & 1.62045\, -0.7110170 i & 1.86066\, -0.7094180 i & 2.10006\, -0.7083100 i & 2.33888\, -0.7075110 i \\
		\noalign{\smallskip}\hline
	\end{tabular}
\end{table*}

\begin{table*}
    \centering
	\caption{Massless scalar quasinormal modes obtained by using the analytic expression \eqref{wanaliticscalar}.}
	\label{tab:3}       
	\begin{tabular}{c|c|c|c|c|c|c}
		\hline\noalign{\smallskip}
		$\epsilon$  & $n$ & $\ell=6$ & $\ell=7$ & $\ell=8$ & $\ell=9$ & $\ell=10$  \\
		\noalign{\smallskip}\hline\noalign{\smallskip}
          \noalign{\smallskip}\hline
        0.0 & 0 & 1.248184\,-0.096225 i & 1.441000\,-0.096225 i & 1.633729\,-0.096225 i & 1.826400\,-0.096225 i & 2.019029\,-0.096225 i \\
      	\noalign{\smallskip}\hline    
        0.1 & 0 & 1.269623\,-0.096758 i & 1.465722\,-0.096758 i & 1.661736\,-0.096758 i & 1.857693\,-0.096758 i & 2.053608\,-0.096758 i \\
      	\noalign{\smallskip}\hline
        0.2 & 0 & 1.292202\,-0.097285 i & 1.491758\,-0.097285 i & 1.691232\,-0.097285 i & 1.890649\,-0.097285 i & 2.090026\,-0.097285 i \\
       	\noalign{\smallskip}\hline
        0.3 & 0 & 1.316015\,-0.097805 i & 1.519218\,-0.097805 i & 1.722340\,-0.097805 i & 1.925407\,-0.097805 i & 2.128435\,-0.097805 i \\
     	\noalign{\smallskip}\hline
        0.4 & 0 & 1.341158\,-0.098315 i & 1.548212\,-0.098315 i & 1.755186\,-0.098315 i & 1.962106\,-0.098315 i & 2.168988\,-0.098315 i \\
       	\noalign{\smallskip}\hline
        0.5 & 0 & 1.367726\,-0.098812 i & 1.578847\,-0.098812 i & 1.789892\,-0.098812 i & 2.000884\,-0.098812 i & 2.211839\,-0.098812 i \\
       	\noalign{\smallskip}\hline
       \noalign{\smallskip}\hline
	\end{tabular}
\end{table*}

\begin{table*}
    \centering
	\caption{Massless Dirac quasinormal modes obtained by using the analytic expression \eqref{eikonal-Dirac}.}
	\label{tab:4}       
	\begin{tabular}{c|c|c|c|c|c|c}
		\hline\noalign{\smallskip}
		$\epsilon$  & $n$ & $\xi=6$ & $\xi=7$ & $\xi=8$ & $\xi=9$ & $\xi=10$  \\
		\noalign{\smallskip}\hline\noalign{\smallskip}
          \noalign{\smallskip}\hline
        0.0 & 0 & 1.149104\,-0.096225 i & 1.342343\,-0.096225 i & 1.535387\,-0.096225 i & 1.728302\,-0.096225 i & 1.921125\,-0.096225 i \\
      	\noalign{\smallskip}\hline    
        0.1 & 0 & 1.168390\,-0.096760 i & 1.364830\,-0.096760 i & 1.561078\,-0.096760 i & 1.757197\,-0.096760 i & 1.953224\,-0.096760 i \\
      	\noalign{\smallskip}\hline
        0.2 & 0 & 1.187676\,-0.097294 i & 1.387318\,-0.097294 i & 1.586769\,-0.097294 i & 1.786092\,-0.097294 i & 1.985324\,-0.097294 i \\
       	\noalign{\smallskip}\hline
        0.3 & 0 & 1.206962\,-0.097829 i & 1.409806\,-0.097829 i & 1.612460\,-0.097829 i & 1.814987\,-0.097829 i & 2.017424\,-0.097829 i \\
     	\noalign{\smallskip}\hline
        0.4 & 0 & 1.226248\,-0.098363 i & 1.432293\,-0.098363 i & 1.638151\,-0.098363 i & 1.843882\,-0.098363 i & 2.049523\,-0.098363 i \\
       	\noalign{\smallskip}\hline
        0.5 & 0 & 1.245534\,-0.098898 i & 1.454781\,-0.098898 i & 1.663841\,-0.098898 i & 1.872777\,-0.098898 i & 2.081623\,-0.098898 i \\
       	\noalign{\smallskip}\hline
       \noalign{\smallskip}\hline
	\end{tabular}
\end{table*}

\begin{table*}
    \centering
	\caption{Relative Error for massless scalar quasinormal modes obtained by using the expressions \eqref{ErrorR}-\eqref{ErrorI} and \eqref{Error}.}
	\label{tab:5}       
	\begin{tabular}{c|c|c|c|c|c|c}
		\hline\noalign{\smallskip}
		$\epsilon$  & $n$ & $\delta_R(\omega(\ell=6))$ & $\delta_R(\omega(\ell=7))$ & $\delta_R(\omega(\ell=8))$ & $\delta_R(\omega(\ell=9))$ & $\delta_R(\omega(\ell=10))$  \\
		\noalign{\smallskip}\hline\noalign{\smallskip}
        \noalign{\smallskip}\hline
0.0 & 0 & (0.296623, 0.0832381) & (0.222654, 0.0625485) & (0.173272, 0.048712) & (0.138672, 0.0390054) & (0.113491, 0.031935)  \\   
      	\noalign{\smallskip}\hline    
0.1 & 0 & (0.289939, 0.0812636) & (0.217652, 0.0610567) & (0.169391, 0.047544) & (0.135576, 0.0380645) & (0.110966, 0.031159)  \\
      	\noalign{\smallskip}\hline
0.2 & 0 & (0.283642, 0.0788558) & (0.213096, 0.0591477) & (0.165996, 0.045969) & (0.132994, 0.0367240) & (0.108974, 0.029990) \\
       	\noalign{\smallskip}\hline
0.3 & 0 & (0.279399, 0.0749502) & (0.210655, 0.0557582) & (0.164758, 0.042925) & (0.132596, 0.0339220) & (0.109189, 0.027365) \\
     	\noalign{\smallskip}\hline
0.4 & 0 & (0.280154, 0.0675349) & (0.213279, 0.0488777) & (0.168626, 0.036402) & (0.137336, 0.0276499) & (0.114563, 0.021275) \\
       	\noalign{\smallskip}\hline
0.5 & 0 & (0.290315, 0.0533578) & (0.225373, 0.0352559) & (0.182010, 0.023151) & (0.151623, 0.0146590) & (0.129506, 0.008474) \\
       	\noalign{\smallskip}\hline
       \noalign{\smallskip}\hline
	\end{tabular}
\end{table*}

\begin{table*}
    \centering
	\caption{Relative Error for massless Dirac quasinormal modes obtained by using the expressions \eqref{ErrorR}-\eqref{ErrorI} and \eqref{Error}.}
	\label{tab:6}       
	\begin{tabular}{c|c|c|c|c|c|c}
		\hline\noalign{\smallskip}
		$\epsilon$  & $n$ & $\delta_R(\omega(\xi=6))$ & $\delta_R(\omega(\xi=7))$ & $\delta_R(\omega(\xi=8))$ & $\delta_R(\omega(\xi=9))$ & $\delta_R(\omega(\xi=10))$  \\
		\noalign{\smallskip}\hline\noalign{\smallskip}
        \noalign{\smallskip}\hline
0.0 & 0 & (0.344967, 0.0207621) & (0.253944, 0.0152366) & (0.194659, 0.011659) & (0.153925, 0.0092105) & (0.124746, 0.007459) \\   
      	\noalign{\smallskip}\hline    
0.1 & 0 & (0.379918, 0.0182290) & (0.290832, 0.0128585) & (0.232806, 0.009381) & (0.192936, 0.0069990) & (0.164376, 0.005296) \\
      	\noalign{\smallskip}\hline
0.2 & 0 & (0.502863, 0.0098912) & (0.415511, 0.0046792) & (0.358612, 0.001303) & (0.319515, 0.0010107) & (0.291507, 0.002665) \\
       	\noalign{\smallskip}\hline
0.3 & 0 & (0.718556, 0.0076027) & (0.632737, 0.0126523) & (0.576835, 0.015925) & (0.538421, 0.0181683) & (0.510904, 0.019772) \\
     	\noalign{\smallskip}\hline
0.4 & 0 & (1.032880, 0.0385003) & (0.948394, 0.0433823) & (0.893359, 0.046549) & (0.855541, 0.0487192) & (0.828449, 0.050272) \\
       	\noalign{\smallskip}\hline
0.5 & 0 & (1.453080, 0.0882365) & (1.369730, 0.0929443) & (1.315430, 0.095999) & (1.278120, 0.0980950) & (1.251390, 0.099594) \\
       	\noalign{\smallskip}\hline
       \noalign{\smallskip}\hline
	\end{tabular}
\end{table*}



%

\bibliographystyle{unsrt}
\bibliography{biblioIL2.bib}

%

\end{document}